\RequirePackage{filecontents}

\documentclass[a4paper,fleqn,usenatbib,useAMS]{mnras}

\usepackage{graphicx}	
\usepackage{amsmath}	
\usepackage{amssymb}	

\usepackage{multicol}        
\usepackage{bm}		
\usepackage{pdflscape}	
\usepackage[table,xcdraw]{xcolor}

\usepackage[T1]{fontenc}
\usepackage{ae,aecompl}

\usepackage{multirow}

\usepackage{newtxtext,newtxmath}
\newcommand{\mass}{{\mathcal{M}}}
\newcommand{\ssfr}{{\mathcal{S}}}

\newcommand{\new}[1]{#1} 
\DeclareMathOperator\erf{erf}

\title[The GSMF \& LSBGs from CCSNe]{The Galaxy Stellar Mass Function and Low Surface Brightness Galaxies from Core-Collapse Supernovae}

\author[T. M. Sedgwick]{Thomas M. Sedgwick$^{1}$\thanks{Contact e-mail: \href{mailto:T.M.Sedgwick@2013.ljmu.ac.uk}{T.M.Sedgwick@2013.ljmu.ac.uk}},
Ivan K. Baldry$^{1}$, Philip A. James$^{1}$ \& Lee S. Kelvin$^{1}$
\\
$^{1}$Astrophysics Research Institute, Liverpool John Moores University, IC2, Liverpool Science Park, 146 Brownlow Hill, L3 5RF
}

\date{Accepted by MNRAS, 2019 January.}

\pubyear{2019}

\begin{document}
\label{firstpage}
\pagerange{\pageref{firstpage}--\pageref{lastpage}}
\maketitle

\begin{abstract}
We introduce a method for producing a galaxy sample unbiased by surface brightness and stellar mass, by selecting star-forming galaxies via the positions of core-collapse supernovae (CCSNe). Whilst matching $\sim$2400 supernovae from the SDSS-II Supernova Survey to their host galaxies using IAC Stripe 82 legacy coadded imaging, we find $\sim$150 previously unidentified low surface brightness galaxies (LSBGs). Using a sub-sample of $\sim$900 CCSNe, we infer CCSN-rate and star-formation rate densities as a function of galaxy stellar mass, and the star-forming galaxy stellar mass function. Resultant star-forming galaxy number densities are found to increase following a power-law down to our low mass limit of $\sim10^{6.4}$ M$_{\odot}$ by a single Schechter function with a faint-end slope of $\alpha = -1.41$. Number densities are consistent with those found by the EAGLE simulations invoking a $\Lambda$-CDM cosmology. Overcoming surface brightness and stellar mass biases is important for assessment of the sub-structure problem. In order to estimate galaxy stellar masses, a new code for the calculation of galaxy photometric redshifts, zMedIC, is also presented, and shown to be particularly useful for small samples of galaxies.
\end{abstract}

\begin{keywords}
galaxies: luminosity function, mass function --- galaxies: star formation --- galaxies: distances and redshifts --- supernovae: general --- methods: statistical
\end{keywords}

\begingroup
\let\clearpage\relax
\endgroup
\newpage

\section{Introduction}\label{sec:intro}
The galaxy stellar mass function (GSMF) is a direct probe of galaxy evolution, as mass is known to be a primary driver of differences in galaxy evolution. For example, \citet{KAU03} find galaxy colours, star-formation rates and internal structure all correlate strongly with stellar mass. It is argued by \citet{THO10} that early-type galaxy formation is independent of environment and controlled solely by self-regulation processes, which depend only on intrinsic galaxy properties including mass. \citet{PAS09} demonstrate that star-formation and AGN activity show the strongest correlations with stellar mass. Past attempts to measure the low-redshift GSMF have established clear evidence of a low-mass upturn in galaxy counts, indicating that low mass galaxies dominate the galaxy population by number at current epochs (\citealt{B12}, henceforth B12; see also \citealt{COL01,BEL03,BAL08,LW09,KEL14b})

The majority of cosmological simulations today invoke a $\Lambda$-CDM description of our universe, due to its ability to simultaneously reproduce various observable properties of the universe \citep{PER99,BEN13}. Despite these successes, a major challenge to the $\Lambda$-CDM model today is the \textquotesingle sub-structure problem\textquotesingle. Numbers of dwarf galaxies as predicted by straightforward simulations are significantly larger than those observed, and as a consequence, so too is the overall number of galaxies on cosmological scales \citep{MOO99}. This discrepancy in dwarf galaxy counts is reflected in the form of the GSMF.

The observed number density of dwarf galaxies increases down to $\sim 10^8$ M$_{\odot}$, below which the form of the GSMF is uncertain (B12). Cosmological simulations of galaxy evolution such as EAGLE \citep{SCH15,CRA15} and ILLUSTRIS \citep{GEN14} are now sophisticated enough to attempt to assess the GSMF into the dwarf regime. In contrast to observational results, the GSMF from these simulations sees number densities continue to increase in the dwarf mass regime, approximately following a power-law.

This discrepancy may not be a fault of a $\Lambda$-CDM description of our universe, however: almost all galaxy surveys suffer from a combination of magnitude and surface brightness constraints \citep{CD02,WRI17}. Most dwarf systems (typically $\lesssim$10$^{8}$M$_{\odot}$, \citealt{KIR13}) have intrinsically lower surface brightnesses than their higher mass counterparts, and consequently, the lower mass end of the GSMF may be underestimated due to sample incompleteness, with lower surface brightness galaxies more likely to be missed by galaxy surveys. The current observed number densities of low mass galaxies can be treated as a lower limit when constraining evolutionary models \citep{BAL08}.

Knowing the precise form of the GSMF is clearly crucial should we wish to use it as a diagnostic of galaxy evolution. Developing techniques to increase completeness of the low mass end of the GSMF must be the focus should one wish to use it to assess the nature of the physics which controls this evolution.

Here we develop and implement one such technique, using the Stripe 82 Supernova Survey \citep[][henceforth, S18]{S18} to produce a sample of galaxies located at the positions of core-collapse supernovae (CCSNe). As CCSNe peak at luminosities of $10^8$ -- $10^9$ $L_{\odot}$, they can be used as pointers to their host galaxies, which may have been missed from previous galaxy surveys due to their low surface brightness: The Palomar Transient Factory \citep{LAW09} located low surface brightness galaxies (LSBGs) when combining SN positions with imaging taken pre-supernova or long after SN peak epoch \citep{PER16}. As well as aiding the identification of LSBGs, a galaxy selection using a complete sample of supernovae may significantly reduce surface brightness and magnitude biases if the host galaxy is identified for each SN in a sample.

Given that the progenitor stars of core-collapse supernovae (CCSNe henceforth) have high masses, it is natural to use them as a tracer of recent star formation. The lower mass limit for zero-age main sequence stars that end their lives as CCSNe has been closely constrained by numerous studies, with the review of \citet{SMA09} presenting a consensus value of 8$\pm$1\,M$_{\odot}$. The upper-mass limit is much more uncertain, due to the possibility that the highest mass stars may collapse directly to black holes, with no visible explosion. However, it seems likely that stars at least as massive as 30\,M$_{\odot}$ explode as luminous CCSNe; \citet{BOT17} adopt an upper mass limit of 40\,M$_{\odot}$. The corresponding range of lifetimes of CCSN progenitors is then something like 6 -- 40\,Myr, for single star progenitors \citep[see, e.g.][]{MAU17}; mass-exchange in high-mass binary stars can extend these lifetimes \citep[e.g.][]{ST15}. Even with this extension, it is clear that on the timescales relevant for studies of galaxy evolution, rates of CCSNe can be taken as a direct and virtually instantaneous tracer of the current rate of star formation.

Several studies have made use of CCSNe as an indicator of star formation in the local Universe. On the most local scales, both \citet{BOT12} and \citet{XE15} have compared CCSN rates and integrated star formation rates within a spherical volume of radius 11\,Mpc centred on the Milky Way, finding good agreement between observed and predicted numbers of SNe. A similar conclusion was also reached by \citet{CET99}, looking at a rather more extended (mean distance $\sim$40\,Mpc) sample of SNe and host galaxies. Other studies have used CCSNe to probe star formation at intermediate redshifts, e.g. \citet{DAH04} who investigated the increase in the cosmic star-formation rate out to redshift $\sim$0.7, and \citet{BOT17} whose sample of 50 SNe mainly occurred in host galaxies in the redshift range 0.3 -- 1.0. Pushing to still higher redshifts, \citet{STR15} have investigated the cosmic SF history out to z$\sim$2.5 using CCSNe within galaxies from the CANDELS \citep{GRO11} and CLASH \citep{POS12} surveys. 

Supernovae have also been used to investigate SF in different environments and types of galaxies, e.g. in starbursts \citep{MIL13} and galaxies with Active Galactic Nuclei \citep{WDW10}, and to determine the metallicity dependence of the local SF rate \citep{STO13}.

By selecting galaxies using CCSNe and measuring galaxy stellar masses, the resultant number densities as a function of mass imply CCSN-rate densities as a function of mass ($\rho_{CCSN}$) in units of yr$^{-1}$Mpc$^{-3}$, under the assumption that the CCSN sample itself is complete. By assuming a relationship between core-collapse supernova rate and star-formation rate, we are able to trace star-formation rate densities (SFRD) (M$_{\odot}$yr$^{-1}$Mpc$^{-3}$). The well-established star-forming galaxy main sequence \citep{NOE07a,DAV16,MCG17,PEA18} can then be used to determine typical star-formation levels expected for a given stellar mass, to infer star-forming galaxy number densities (Mpc$^{-3}$) as a function of galaxy stellar mass (the GSMF), such that $\rho_{CCSN} \rightarrow \rho_{SFR} \rightarrow$ GSMF. 

A programme similar to the present work was proposed by \citet{CB15} who suggested that SNe detected by the Large Synoptic Survey Telescope from 2021 could be used as a statistical probe of the numbers and stellar masses of dwarf galaxies. The present work can be seen as a precursor to such a study.

The structure of the present work is as follows. Section \ref{sec:eqsbreakdown} outlines in further detail the connections between CCSN-rate density, star-formation rate density, and the galaxy stellar mass function, along with the assumptions required to form them. In Section \ref{sec:Data} we present the data sets used. In Section \ref{sec:Methodology} we outline our methodology for drawing from these complete SN and galaxy samples, unbiased by magnitude and surface brightness, as well as our procedure for obtaining photometric redshift estimates. Section \ref{sec:ResultsandDiscussion} presents resultant SFRD and star-forming GSMF estimates obtained via a CCSN host galaxy selection, where comparison is drawn with existing SFRD and GSMF results, both observational and simulated.

\section{Converting CCSN-Rate Density to the Star-Forming Galaxy Stellar Mass Function}\label{sec:eqsbreakdown}

In this section, we represent mathematically the CCSN-rate density, the SFRD, and the GSMF. This is in order to define the connections between them, and hence, how we are able to arrive at an estimate for the GSMF by measuring the CCSN-rate density as a function of host galaxy stellar mass ($\mass$).

For a volume-limited sample of galaxies, the binned GSMF is defined by 
\begin{equation}\label{eq:GSMF}
  \Phi(\mass) \: = \: \frac{1}{\Delta \log \mass} \: \frac{N(\mass)}{V}    
\end{equation}
over a mass bin of width $\Delta \log \mass$,  where $N$ is the number of galaxies in the bin, and
$V$ is the volume. In other words, the GSMF is the number of galaxies, per unit volume, per
logarithmic bin of galaxy stellar mass.

The SFRD is often estimated for the entire galaxy population, particularly as a function of redshift \citep{MD14}, but it can also be determined as a function of galaxy mass \citep{G10}. This can then be given by
\begin{equation}\label{eq:SFRD2}
  \rho_{\rm SFR}(\mass) \: = \: \frac{1}{\Delta \log \mass} \: \frac{\sum_{i=1}^{N} \, \ssfr_{i} \, \mass_{i}}{V}  
\end{equation}
where $\ssfr_i$ is the {\em specific} star-formation rate (SSFR) for each of the $N$ galaxies in a
bin. In other words, the SFRD is the summed star-formation rate, per unit volume, 
per logarithmic bin of stellar mass.

We can approximate the SFRD by considering that the majority of star formation in the universe
occurs on the galaxy main sequence \citep{NOE07a,DAV16,MCG17}. This sequence represents the relation, and its
scatter, of SFR versus mass for typical star-forming galaxies. The SFRD can then be given by
\begin{equation}\label{eq:SFRD3}
  \rho_{\rm SFR}(\mass) \: = \: \frac{1}{\Delta \log \mass} \: \frac{\overline{\ssfr}(\mass) \: \mass \: N_{\rm SF}(\mass)}{V}   
\end{equation}
where $N_{\rm SF}$ is number of star-forming galaxies in the bin, $\mass$ is the mid-point mass
(assuming ${\Delta \log \mass} \ll 1$), and $\overline{\ssfr}$ is the mean SSFR for star-forming
galaxies.

By star-forming galaxy, we mean all galaxies that are not 
permanently quenched or virtually quenched with minimal 
residual star formation. In other words, these are the galaxies that are represented by the cosmological SFRD as a function of mass. 
Note that in the estimate of the mean $\ssfr$, we should include galaxies that are in a quiescent phase but are otherwise representative 
of the typical star-forming population, and our CCSN-host galaxy selection method naturally leads to an appropriate contribution from such galaxies. This is relevant for low-mass galaxies that undergo more variation in their SFR with time \citep[see, e.g.][]{SKI05,STI07}. The mean $\ssfr$ should represent an average over duty cycles
in this regime.

Comparing with Eq.~\ref{eq:GSMF}, we note that the SFRD can then be rewritten in terms of the GSMF
of star-forming galaxies $\Phi_{\rm SF}$ as follows
\begin{equation}\label{eq:SFRD4}
  \rho_{\rm SFR}(\mass) \: = \: \overline{\ssfr}(\mass) \: \mass \: \Phi_{\rm SF}(\mass)    
\end{equation}
By using a parameterisation of SSFR with galaxy stellar mass, \citep[e.g.][]{NOE07a,SPE14}, it is possible to 
estimate the GSMF for star-forming galaxies from the SFRD or vice versa.  

Next we consider the observed CCSN-rate density. Here we define this as the rate of CCSNe observed
over a defined volume of space (redshift and solid angle limited), per unit volume, per logarithmic bin of
galaxy stellar mass. From a non-targeted supernova survey, like that of SDSS \citep[][S18]{FRI08}, this is given by
\begin{equation}\label{eq:CCSNobs}
  \rho_{\rm CCSN,obs}(\mass) \:  = \: \frac{1}{\Delta \log \mass}  \: \frac{n_{\rm CCSN,obs}(\mass)}{\tau \, V}    
\end{equation}
where $n_{\rm CCSN,obs}$ is the number of observed CCSNe associated with galaxies in the bin, and
$\tau$ is the effective rest-frame time over which CCSNe could be identified.  The time period of the supernova survey, 
in the average frame of the host galaxies ($\tau$), is shorter than that in the observed frame (t), such that $\tau$ = t / (1+$\overline{z}$).

The relationship between the CCSN rate and SFRD is then given by 
\begin{equation}\label{eq:CCSNSFRD}
  \rho_{\rm CCSN,obs}(\mass) \: = \: \rho_{\rm SFR}(\mass) \: \overline{\epsilon}(\mass) \: \overline{\mathcal{R}}(\mass) 
\end{equation}
where $\overline{\mathcal{R}}$ is the mean ratio of CCSN rate to SFR, 
which is equivalent to the number of core-collapse supernovae per mass of stars formed;
and $\overline{\epsilon}$ is the mean efficiency of detecting supernovae.  
For an apparent-magnitude limited supernova survey, the latter function accounts for
varying brightnesses and types of supernova occurring in star-forming galaxies of a given stellar
mass, and the variation in extinction along different lines of sight to the supernovae.

By combining these relations we arrive at 
\begin{equation}\label{eq:CCSNGSMF}
\Phi_{\rm SF} \: = \: \frac{\rho_{\rm CCSN,obs}}{\overline{\epsilon} \: \overline{\mathcal{R}} \: \overline{\ssfr} \: \mass}
\end{equation}
which explicitly relates CCSN-rate density to the star-forming GSMF. The connection 
is given in terms of three functions of galaxy stellar mass: 
$\overline{\ssfr}$ is the SSFR relation of the galaxy main sequence; 
$\overline{\mathcal{R}}$ is the number of CCSNe per unit mass of stars formed; 
and $\overline{\epsilon}$ is the efficiency of detecting CCSNe, 
which depends on their luminosity function, and also non-intrinsic 
effects of sample selection and survey strategy, in particular, 
the limiting CCSN detection magnitude. 
The basic premise is that these should be a weak function of galaxy stellar mass. 
We investigate the effects of varying $\epsilon$
on estimated CCSN-rate densities in Section~\ref{sec:Corrections}, and of varying $\overline{\ssfr}$ on the GSMF in Section~\ref{sec:GSMF}.

\section{Data} \label{sec:Data}

The present study makes use of 3 data sets, all of which are data products of the Sloan Digital Sky Survey (SDSS). SDSS is a large-area imaging survey of mainly the north Galactic cap, with spectroscopy of $\sim$10$^6$ galaxies and stars, and $\sim$10$^5$ quasars \citep{YOR00}. The survey uses a dedicated, wide-field, 2.5 m telescope \citep{GUN06} at Apache Point Observatory, New Mexico. A 142 megapixel camera, using a drift-scan mode \citep{GUN06}, gathers data in 5 optical Sloan broad band filters, \textit{ugriz}, approximately spanning the range from 3000 to 10,000 \AA, on nights of good seeing. Images are processed using the software of \citet{LUP01} and \citet{STO02}. Astrometric calibrations are achieved by \citet{PIE03}. Photometric calibrations are achieved using methods described by \citet{HOG01} and \citet{TUC06} via observations of primary standard stars observed on a neighbouring 0.5m telescope \citep{SMI02}.

We make use of data associated with the Stripe 82 Region, a 275 sq. degree equatorial region of sky \citep{BAL05}. The region spans roughly 20\textsuperscript{h} < R.A. < 4\textsuperscript{h} and --1.26$^{\circ}$ < Decl. < 1.26$^{\circ}$. Between 1998 and 2004, the region was scanned $\sim$80 times. A further $\sim$200 images were taken between 2005 and 2007, as part of the SDSS-II Supernova Survey \citep[][S18]{FRI08}.

The SDSS-II Supernova Survey data release outlined in S18 forms the basis of the supernova sample used for this study. 10258 transient sources were identified using repeat \textit{ugriz} imaging of the region, with light curves and follow-up spectra used for transient classifications, all of which are utilised in this study to produce a SN sample, with great care taken to ensure its completeness and the removal of non-SN transients.

We aim to produce a galaxy sample selected via the SNe which they host. Host galaxies for many of the transient sources were already identified as part of the Supernova Survey. However, in the present study we revisit host-galaxy identification for 2 reasons: i) There is now access to deeper, coadded SDSS imaging with which to search for the host galaxy. ii) There is often a natural bias towards assigning a transient to a higher surface brightness galaxy when one or more lower surface brightness galaxy is nearby. Our method of transient-galaxy matching is designed specifically to address this bias.

To form this galaxy sample, we make use of both single epoch imaging and multiple epoch SDSS imaging. Single epoch imaging published as part of SDSS-IV DR14 forms our initial galaxy sample (referred in the present study as the SDSS sample for simplicity), and the sample of stars used for the removal of variable stars from our SN sample, as outlined in Sections \ref{sec:starremoval} and \ref{sec:SDSScatalogue} respectively. Galaxy and star classification is described in Section 4.4.6 of \citet{STO02}. 

We then turn to coadds of multiple epoch imaging. The IAC Stripe 82 legacy project \citep{FT16} performs median stacking of existing legacy Stripe 82 data, with additional complex sky-subtraction routines applied thereafter, in order to reach the extremely faint limits of the data ($\sim$28.5 mag arcsec$^{-2}$ to 3$\sigma$ for 10$\times$10 arcsec$^2$). The IAC Stripe 82 legacy catalogue hence forms a deeper sample of objects used for this investigation. Approximately 100 single epoch images are median stacked per SN region, to produce the deeper imaging crucial for LSBG detection. From this coadded imaging we aim to identify additional low-surface brightness galaxies not found by the SDSS sample. IAC Stripe 82 image mosaicking and postage stamp creation about the positions of our SNe, crucial for host galaxy identifications, were completed using the Cutout and Mosaicking Tool, part of the ARI Survey Imaging Tools, created by one of the authors (LSK). We compare the completeness of the SDSS sample with the sample found by \citet{FT16} from this coadded data, as well with a \texttt{SExtractor} implementation designed as a bespoke search for CCSN host galaxies, using the same data (Section \ref{sec:sExtractor}), in order to demonstrate the sensitivity of results to sample incompleteness.

We also require redshift estimates for our SN-galaxy pairs. Approximately 480 of the SN candidates have spectra of their own, from ten sources outlined in \citet{FRI08}. We also use host-galaxy spectra for our SN-galaxy pairs once the host galaxy has been confidently identified. The galaxy spectra utilised stem from 3 main sources within SDSS. These are the SDSS-II Legacy \citep{YOR00}, SDSS-II Southern \citep{BAL05}, and SDSS-III BOSS/SDSS-IV eBOSS surveys \citep{DAW13,DAW16}. The latter contains spectroscopy for galaxies identified as the hosts of 3743 of the 10258 SN candidates in S18, approximately a third of which were identified as non-supernovae as a result. Supernova redshifts are used in cases where both are available. Photometric redshifts of galaxies are calculated from the coadded photometry in cases where no spectroscopic redshift is available for a SN-galaxy pair, as outlined in Section \ref{sec:zMedIC}.

\section{Methodology} \label{sec:Methodology}

\subsection{Selection of the Supernova Sample}

\subsubsection{Star Removal}\label{sec:starremoval}

In order to produce a sample of core-collapse supernovae (CCSNe), we first focus on removing non-supernovae from the SDSS-II Supernova Survey sample. 10258 transient sources were found by S18. We build on their classification attempts by firstly removing those transient sources categorised as variable stars (objects detected over multiple observing seasons) and AGN (identified spectroscopically via their broad hydrogen lines).

The main SN classifications of S18 are type II, type Ib/c, and type Ia SNe. We ultimately wish to remove type Ia SNe to obtain a CCSN sample in order to trace star formation rates. However, we keep likely type Ia SNe in the sample at this stage to search for LSBGs and to increase the size of our training sample used for the estimations of galaxy redshifts, as described in Section \ref{sec:zMedIC}. At this stage, the sample consists of 6127 transients. Of these objects, 1809 are spectroscopically confirmed SNe and a further 2305 are, photometrically, deemed very likely to be supernovae, via a combination of Bayesian, nearest-neighbour and light-curve fit probabilities (see S18 for a full description). Those remaining are classified as \textquotesingle Unknown\textquotesingle. However, these objects may still be supernovae. For several of these objects it may simply be unclear from the photometry what type of supernova is being seen. For instance, if probabilities derived from the 3 aforementioned techniques give a reasonable likelihood for more than 1 of type Ia, Ib/c or II, the object will be classified as \textquotesingle Unknown\textquotesingle.

\begin{figure}
	\centerline{\includegraphics[width=1.2\columnwidth]{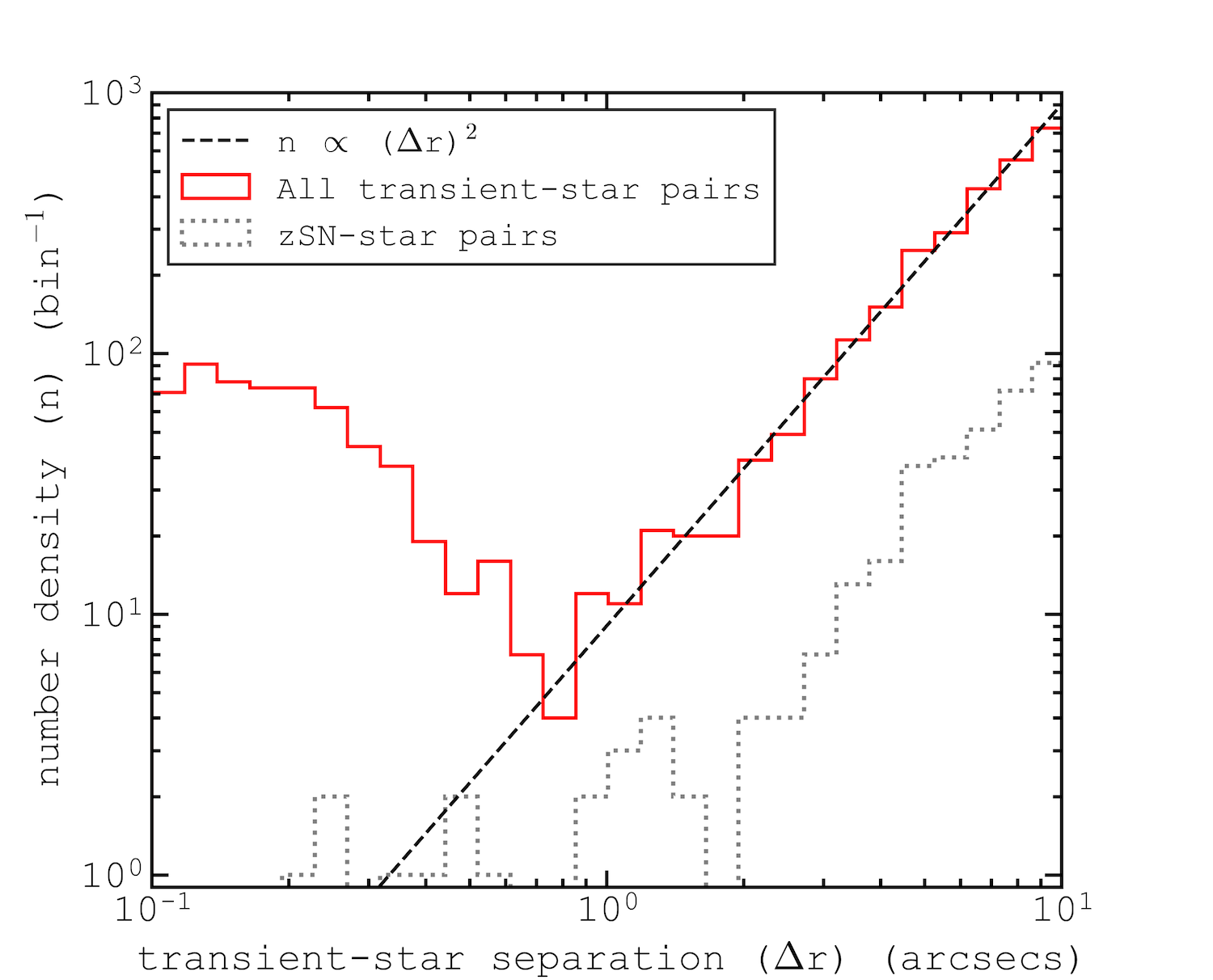}}
    \caption{Number densities of transient--star pairings as a function of transient--star sky separation. The red solid line represents all possible transient--star pairings. The grey dotted line represents all spectroscopically-confirmed-SN--star pairings. The dashed line shows the n = 2 power-law expected to be followed by unassociated transient-star pairs. For a transient--star separation $\lesssim$ 1" transients are likely to be stars.}
    \label{fig:starsep}
\end{figure}

We match transient positions with all objects of the SDSS-IV DR14 PhotoPrimary catalogue located in the Stripe 82 region with Galactic extinction-corrected \textit{r}-band magnitude < 22.0 (Petrosian, psf or model) ($\sim10^7$ objects). We refer to this as the SDSS catalogue in the remainder of the present work, for simplicity. Additional variable stars are found in the SN sample by computing the separations between SDSS stars and all transients without a spectroscopic SN classification. Variable stars are identified as those objects found within 1" of an SDSS star. This 1" transient -- star separation cut-off was chosen following inspection of Figure \ref{fig:starsep}.

The counts of non-associated transient--star pairs rise as the square of their separation. Additional counts arise below a separation of approximately 1" due to genuine association between the transient and SDSS object, and the detection is deemed to have arisen from the star. We remove \new{718} stars from the supernova sample in this manner, leaving a sample of 5549 transients. Most of these transients are likely to be SNe, but some may be QSOs. Redshift information helps distinguish SNe and QSOs. However, not all of these transients have spectroscopic redshifts, as outlined in Section \ref{sec:Data}. We therefore keep all remaining transients in the sample at this stage, until each source is matched to its host galaxy, for which a spectroscopic redshift may be available.

\subsubsection{SN apparent peak magnitude cut}\label{sec:SNmagcut}

With stars removed, we turn our attention to the completeness of the sample. The \textit{r}-band supernova peak-magnitude ($r_{SN,peak}$) distribution is found to follow a power-law up to $\sim21.8$, beyond which the slope decreases. A power-law increase to counts is expected due to the increased volume sampled as mean SN apparent magnitude becomes fainter. A deviation from this power-law for $r_{SN,peak}$ > 21.8 hence indicates incompleteness. We henceforth implement a cut to include only supernovae brighter than $r_{SN,peak} = 21.8$. Approximately 25\% of the aforementioned removed stars are brighter than this cut. Table \ref{tab:SNsamplesizes} shows the SN sample sizes following the removal of stars and spectroscopically confirmed AGN. SN counts are also given as a function of SN type. The $r_{SN,peak} = 21.8$ SN apparent magnitude cut reduces the sample to 2931 SNe.

\begin{table}
\caption{\new{Transient counts as a function of type, built from the SDSS-II Supernova Survey sample of 10258 transients. Counts are divided into: i) those rejected by a magnitude cut, {$r_{SN,peak}$>21.8}, because of sample incompleteness for fainter transients; ii) those brighter than {$r_{SN,peak}$<21.8} but rejected as variable stars or AGN; and iii) those that are selected for our sample. Variable star counts are shown as the summation of those classified by \citet{S18}, by host matching to single-epoch SDSS imaging, and to Stripe 82 legacy coadded imaging. AGN counts are shown as the summation of those classified by \citet{S18}, by host matching to single-epoch SDSS imaging, and from redshifts that indicate the host is a QSO (See Sections \ref{sec:SDSScatalogue} to \ref{sec:sExtractor}). Counts are also subdivided into those classified using spectroscopy and those using photometry.}}
\centering
\resizebox{\columnwidth}{!}{%
\begin{tabular}{c|c|cc|c}
 & transient type & spec & phot & total \\ \hline
\multirow{7}{*}{\begin{tabular}[c]{@{}c@{}}$r_{SN,peak}>21.8$\\ (5257)\end{tabular}} & Ia & 301 & 302 & 603 \\
 & Ib/c & 17 & 12 & 29 \\
 & II & 149 & 813 & 962 \\
 & SL & 0 & 0 & 0 \\
 & Unknown & 0 & 884 & 884 \\
 & Variable Star & 0 & 2416 & 2416 \\
 & AGN & 363 & 0 & 363 \\ \hline
\multirow{2}{*}{\begin{tabular}[c]{@{}c@{}}$r_{SN,peak}<21.8$\\ rejected (2545)\end{tabular}} & Variable Star & 0 & 1342 + 185 + 294 & 1821 \\
 & AGN & 543 + 98 + 83 & 0 & 724 \\ \hline
\multirow{5}{*}{\begin{tabular}[c]{@{}c@{}}$r_{SN,peak}<21.8$\\ selected\\ (2456)\end{tabular}} & Ia & 966 & 267 & 1233 \\
 & Ib/c & 51 & 7 & 58 \\
 & II & 274 & 207 & 481 \\
 & SL & 3 & 0 & 3 \\
 & Unknown & 0 & 681 & 681 \\ \hline
Total & - & 2848 & 7410 & 10258
\end{tabular}%
}
\label{tab:SNsamplesizes}
\end{table}

\subsection{SN Host Galaxy Identification}

\subsubsection{SDSS Catalogue}\label{sec:SDSScatalogue}

Following supernova sample completeness checks, we aim to determine the correct host galaxy for each supernova. Firstly, similar to the aforementioned transient--star matching, for each supernova we find the separation from each SDSS galaxy within a 130" radius. We then size-normalise this separation to be in units of the galaxy\textquotesingle \,s Petrosian radius. To do this, we take the following steps:

\begin{enumerate}
    \item \textbf{Flag unreliable radii:} Galaxy Petrosian radii calculations are deemed reliable if \textit{all} of the following SDSS flag criteria \citep{LUP01} are met: a) \verb|NOPETRO = 0|; b) \verb|petroradErr_r| > 0; c) \verb|clean = 1|; d) \verb|petroR90Err_r|/\verb|petroR90_r| < 1.
    \item \textbf{Winsorize Petrosian Radii:} Winsorization is the limiting of extreme values to reduce the effects of potentially spurious outliers \citep{HAS47}. We set a minimum radius of 2", and if the radius is flagged as unreliable, set a maximum radius of 10". This radius maximum prevents a galaxy with an unphysically large radius measurement from being matched to a distant, unassociated supernova. The radius minimum ensures SN-galaxy matches are not missed due to an underestimation of radius. 2" is chosen as a minimum as it approximates the radius of the lowest stellar mass galaxies known to be in SDSS Stripe 82 at the lowest redshifts in our sample (see Section \ref{sec:zspec}), whereas the maximum of 10" corresponds to the 90th percentile of radius in the SDSS Stripe 82 galaxy sample.
    
    \item \textbf{Account for galaxy eccentricity:} Axis ratio $b/a$ from an exponential fit is Winsorized to $b/a$ > 0.5. From axis ratio and Petrosian radius, we calculate $r_{gal,proj}$: the length on the sky projected from the galaxy centre, to the edge of the galaxy ellipse, in the direction of the supernova.
\end{enumerate}

SN-galaxy separation is then normalised to units of this projected galaxy radius, and for each SN, the 3 galaxies with the lowest normalised separations are taken as the 3 most likely host candidates.
The Petrosian radius is chosen for this method due to the robustness of measurements over a large redshift range \citep{STO02}. To improve confidence in the most likely host galaxy for each SN, we then consider the following 3 factors:

\begin{enumerate}
    \item \textbf{Is the normalised separation reasonable?} A separation of < 1.25 galaxy radii is deemed as a likely association, based on a similar analysis as seen in Figure \ref{fig:starsep}. If no galaxy lies within 1.25 radii of the SNe, the host is flagged as ambiguous. 
    \item \textbf{Are SN and galaxy redshift compatible?} We use the 10th and 90th percentiles of expected SN absolute magnitude, according to \citet[][henceforth, R14]{RIC14}. Different distributions are used for each SN type. We then draw from these and the observed SN apparent magnitude a range of possible redshifts. Should the SN and galaxy redshift appear inconsistent, the host is ambiguous.
    \item \textbf{Is the Petrosian radius reliable?} (see above).
\end{enumerate}	

\begin{figure*}
	\includegraphics[width=0.95\textwidth]{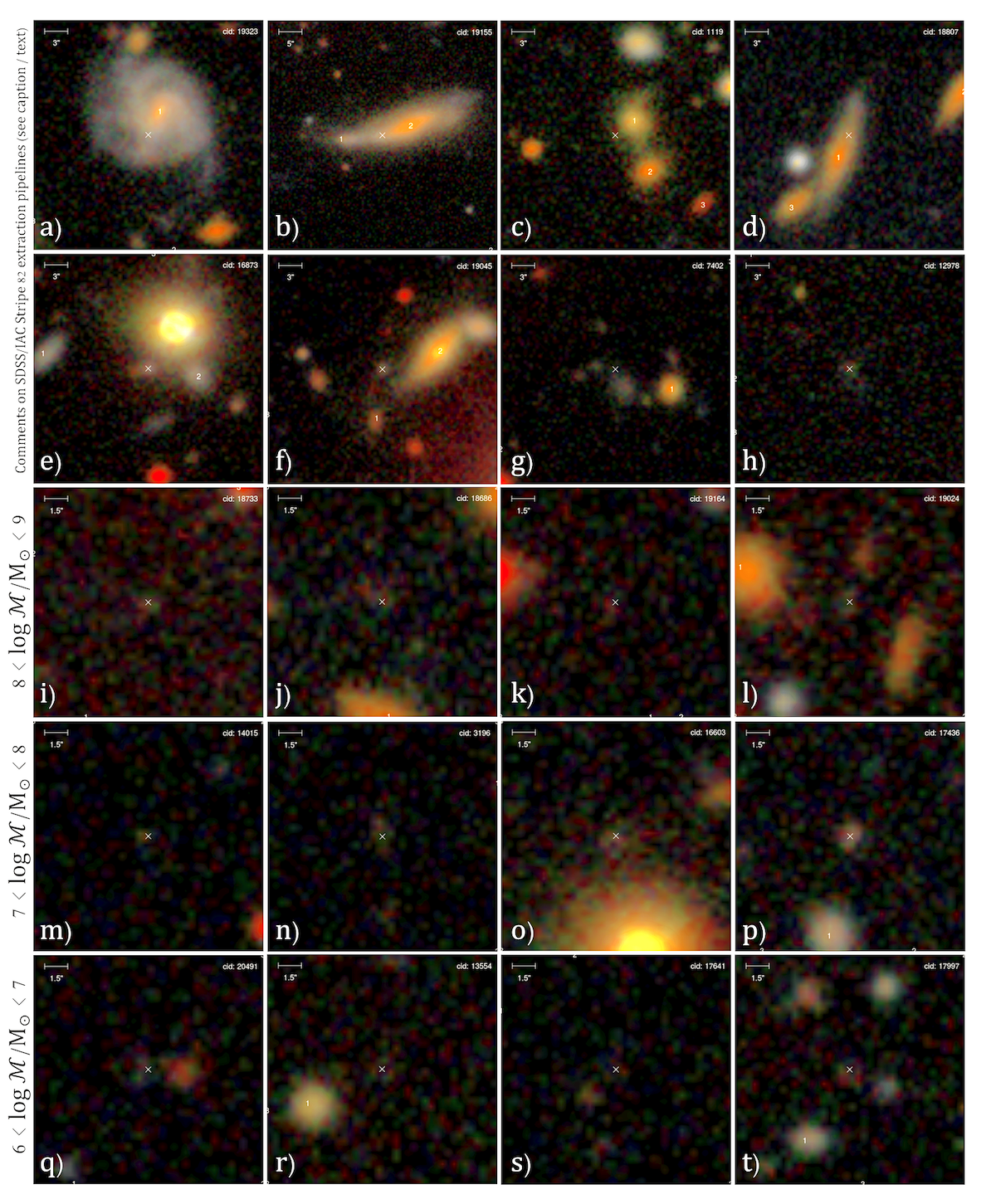}
    \caption{Example postage stamps of IAC Stripe 82 legacy coadded imaging, centred on SN positions, used to inspect each SN region to identify its host galaxy. SN catalogue ID, as listed in the SDSS-II Supernova Survey is indicated. Labels 1, 2 \& 3 in each stamp indicate the central positions of the 3 normalised-nearest SDSS galaxies to the SN. \textbf{Row 1:} chosen to show a) a straightforward SN-galaxy match, b) the successful resolution of satellite galaxies from their primaries, as well as a particularly ambiguous case, \new{c) resolved galaxies in group environments, and d) a case involving extreme morphology. For a) to d), the normalised-nearest galaxy is chosen as the host galaxy}. \textbf{Row 2}: example SNe associated with newly identified LSBGs in IAC imaging. Uncatalogued SN hosts are missed by the SDSS and IAC catalogues due to: g) and h) low-surface-brightness alone; f) being a close-in satellite or the lower-luminosity constituent of a merger; or e) a bright neighbour. \textbf{Rows 3-5:} Examples of newly identified LSBGs. Rows bin by mean galaxy stellar mass from a Monte Carlo technique (see Section \ref{sec:MC}).}
    \label{fig:postagestamps}
\end{figure*}

Each SN region is inspected using IAC Stripe 82 Legacy coadded \textit{gri} imaging \citep{FT16}, with the above flags used to aid the search for a host galaxy. Figure \ref{fig:postagestamps} shows example \textit{gri} imaging used for inspection, with supernova position and host galaxy candidates labelled. At this stage, we only allow SNe to be assigned to galaxies in the SDSS catalogue, such that we can test the performance of the SDSS single-epoch imaging inferred sample for the task of host identification.

Following manual inspection of the coadded \textit{gri} imaging, our procedure finds that 86\% of $r_{SN,peak} < 21.8$ supernovae are matched to an $r < 22.0$ SDSS galaxy. 96\% of these are ultimately assigned to their normalised-nearest galaxy. When deciding the normalised nearest galaxy to each SN without accounting for galaxy ellipticity, this fraction is reduced to $\sim$94\%. Approximately 2\% are matched to the second normalised-nearest. Reasons why the normalised-nearest galaxy is not the SN host include that fact that the galaxy extraction pipeline of SDSS sometimes catalogues a galaxy\textquotesingle s star-forming region or the galaxy bulge as a galaxy in its own right. Another reason comes in cases of \new{extreme galaxy eccentricities or irregular galaxy morphologies}. Only $\sim$0.1\% of SNe are assigned to the third normalised-nearest galaxy. The remaining $\sim$2\% are assigned to another $r < 22.0$ SDSS galaxy outside of the 3 normalised-nearest galaxies. The most common reason for this is again the case of bright galaxies with well-resolved stellar/dust components, and in particular, where a bright bulge was catalogued as the galaxy, and the disk was missed by the extraction pipeline altogether. 

We emphasise the importance of taking the described precautions before assigning SNe to host galaxies, as it is more likely that SNe belonging to dwarf or satellite galaxies are assigned to the wrong host. Matching without caution would almost certainly cause a biased selection towards brighter galaxies which would ultimately reflect in our GSMF.

\new{98 of the SDSS galaxies matched to SN candidates were previously classified QSOs. The AGN were then assumed to be the source of these galaxies' transients. Removing these leaves \new{at this stage} 2422 SN candidates with hosts found in single epoch SDSS imaging.}

\subsubsection{IAC Stripe 82 Legacy Catalogue}

To find the host galaxy for the remaining 14\% of SNe, attention is turned to the deeper IAC Stripe 82 Legacy Catalogue \citep{FT16}, extracted from the same coadded \textit{gri} imaging used for the above manual inspection. Data are formed from the median stacks of $\sim$100 images per supernova region, with individual epoch imaging stemming from both the SDSS-main data release \citep{STO02} and from repeat visits of the supernova regions between 2005 and 2007 as part of the SDSS SN survey (S18).
Both the median stacks and SN peak-magnitude-epoch imaging are inspected to ensure that the point sources are not visible in the former, and are visible in the latter, such that neither supernova nor host galaxy are spurious detections.

\new{The increased signal-to-noise achieved by the coadded imaging showed that 294 of the objects matched to SN candidates which were classified by the single epoch imaging as galaxies were more likely to be stars. The corresponding SN candidates were removed accordingly, leaving \new{at this stage} 2128 galaxies identified from single-epoch imaging in the SN host sample.}

\new{The SDSS and IAC catalogues were linked}, by matching their objects within 2". Removing objects in common, we then repeat the previous matching procedure using Kron-based magnitudes and radii in the place of Petrosian measurements, due to the increased flexibility of Kron-based magnitudes in capturing a large fraction of the object flux \citep{KRO80}.
Note that these objects have photometry only, and do not have redshift estimates. Therefore, we do not consider the possible redshift range, nor the trustworthiness of the radius measurement in our matching procedure.

Nevertheless, $\sim$1\% and \new{$\sim$64\%} of the previously unmatched supernovae candidates are matched to a star or to a galaxy from the IAC catalogue, respectively. This still leaves 146 supernovae not matched to either a galaxy or a star in either catalogue. We hence turn to our own \texttt{SExtractor} procedure, applied to the same coadded imaging, to attempt to locate the remaining host galaxies.

\subsubsection{Bespoke LSBG Identification}\label{sec:sExtractor}

Care is taken to obtain photometry for the remaining host galaxies, whilst simultaneously obtaining photometry for the previously matched objects which is consistent with that found by SDSS and IAC. To do this we consider several \texttt{SExtractor} parameter setups \citep{BER96}, using different detection thresholds, minimum aperture sizes, and smoothing filters, as shown in Table \ref{tab:sextractor}. \textit{ugriz} magnitudes are calculated using a fixed \textit{r}-band defined, elliptical Kron aperture for all bands.

\begin{table*}
\caption{\texttt{SExtractor} configurations used for a bespoke search for LSBGs. Parameters shown are the only parameters varied for each configuration, from the setup of \citet{FT16} (here labelled IAC\_S82). \new{2391} identified galaxies using the setup IAC\_S82 are matched to our SN catalogue. \texttt{SExtractor} configurations are implemented in the order 1) to 6), and photometry from the first configuration to detect the galaxy is used. Additional SN-galaxy matches with successive configurations are indicated.}
\resizebox{\textwidth}{!}{%
\begin{tabular}{l|lllll|l}
\texttt{SExtractor} Configuration & DETECT\_MINAREA (pixels$^2$) & DETECT\_THRESH & FILTER\_NAME & DEBLEND\_NTHRESH & DEBLEND\_MINCONT & Additional SN host-galaxy detections\\ \hline
1) IAC\_S82 & 4 & 2 & default.conv & 16 & 0.002 & 0 \\
2) default\_3.0 & 3 & 1 & default.conv & 32 & 0.001 & 63 \\
3) tophat\_2.5 & 3 & 1 & tophat\_2.5\_3x3.conv & 64 & 0.0001 & 28 \\
4) mexhat\_2.5 & 3 & 1 & mexhat\_2.5\_7x7.conv & 64 & 0.0001 & 23 \\
5) tophat\_1.5 & 3 & 1 & tophat\_1.5\_3x3.conv & 64 & 0.0001 & 18 \\
6) gauss\_1.5 & 2 & 0.5 & gauss\_1.5\_3x3.conv & 64 & 0.0001 & \new{9}
\end{tabular}%
}
\label{tab:sextractor}
\end{table*}

Each of these \texttt{SExtractor} setups is more prone to extracting spurious objects than the last, due to its increasingly generous extraction threshold. However, as we only keep detections corresponding to the likely supernova host, and as the supernova's presence is confirmed by visual inspection, the spurious detections are not deemed problematic. 

To maximise the reliability of our photometry and to minimise the number of these false detections, we first generate a galaxy catalogue using the first \texttt{SExtractor} setup, i.e. the most reliable. This was the setup used by \citet{FT16}, and we successfully reproduce their catalogue. We then complete the aforementioned SN-galaxy matching procedure using this catalogue. If the host galaxy is not found, we turn to the next \texttt{SExtractor} setup, to create a catalogue of the previously missed objects, and so on, for each of our setups, until a galaxy match is found for each SN. In the case of 5 SN, none of our \texttt{SExtractor} setups could automatically detect the galaxy host. In these cases we assume the presence of a host galaxy and a 2.5" circular aperture is used for the galaxy\textquotesingle s photometry, centred on the \textit{r}-band SN position. Indeed, these 5 galaxies, along with their supernovae, may be spurious, but can still be used to assess uncertainties on the form of the GSMF.

Using our procedure, the photometry of the host galaxy is the photometry derived from the \texttt{SExtractor} setup which first located it. The galaxies which required several extraction attempts are therefore subject to the largest uncertainties. However, these uncertainties are folded into our results and can help constrain the form of the galaxy stellar mass function considerably. 

In the case that the object is detected in the \textit{r}-band but is not detected in another band, the magnitude is set to 3 times the sky noise in the aperture for that band. However, we attempt to avoid using these magnitudes wherever possible. Only in the case of 6 galaxies are we forced to use these magnitudes, due to the lack of reliable photometry in other bands.

Row 2 of Figure \ref{fig:postagestamps} shows example IAC Stripe 82 legacy coadded imaging in which the supernova is centred on a newly identified LSBG from our bespoke search. Uncatalogued SN hosts were missed by the SDSS and IAC catalogues due to i) low-surface-brightness alone; ii) a bright neighbour; or iii) being a close-in satellite or the lower-luminosity constituent of a merger.

The different \texttt{SExtractor} setups shown in Table \ref{tab:sextractor} were found to overcome different problems faced by the SDSS/IAC extraction pipelines. For instance, a \textquotesingle Mexican hat\textquotesingle-type smoothing filter was particularly useful for identifying dwarf satellite galaxies close to brighter companions.

\subsubsection{Summary}

For the same reasons given for a galaxy being missed by the SDSS and IAC catalogues, we find that $\sim$2\% of hosts were incorrectly identified by S18. The vast majority of galaxies constituting this 2\% are more massive than the true SN host, and thus their inclusion in our galaxy sample would cause an underestimation of dwarf galaxy counts.

\begin{figure}
	\centerline{\includegraphics[width=1.2\columnwidth]{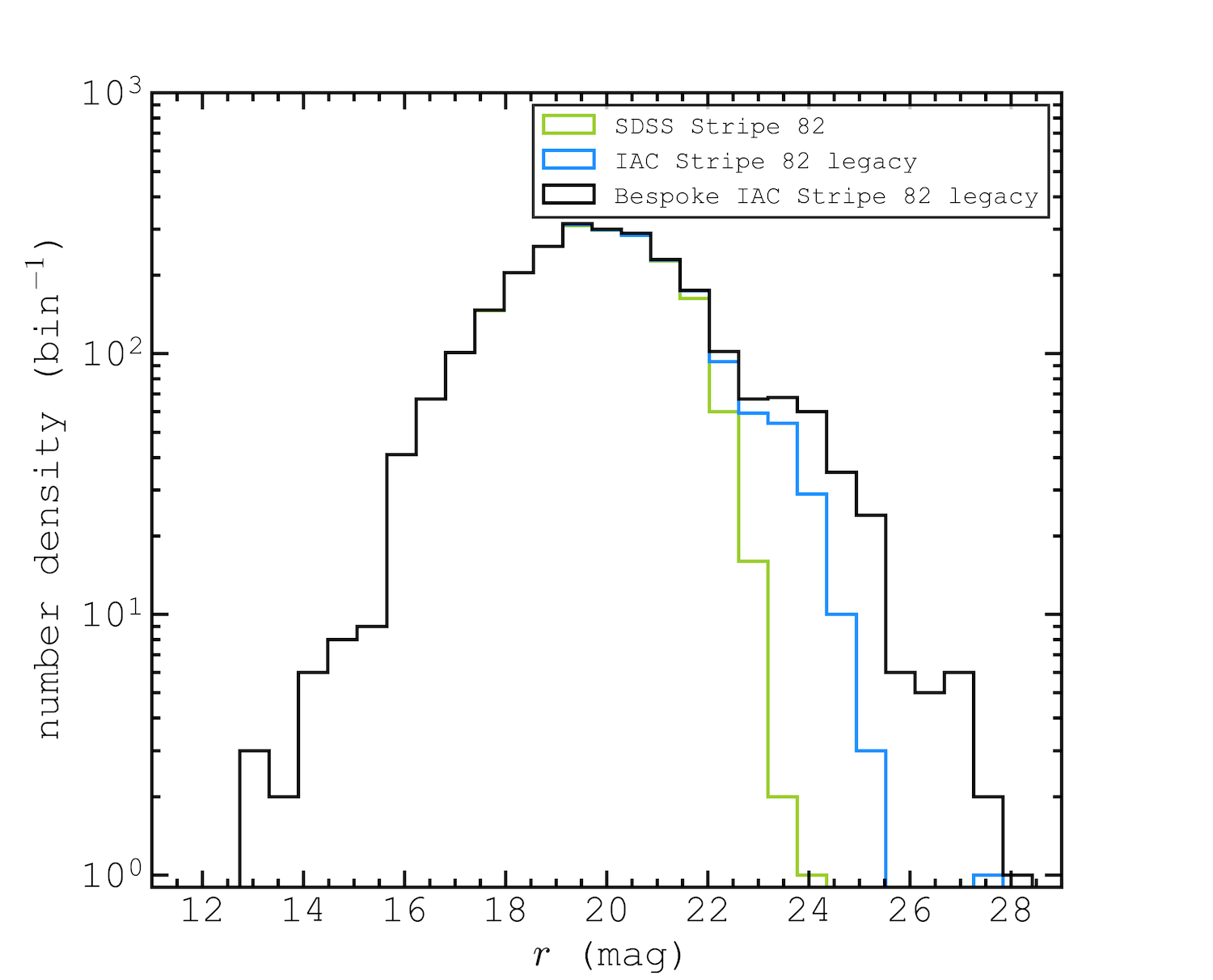}}
    \caption{Galaxy number densities as a function of Kron \textit{r}-band magnitude calculated from a bespoke search of IAC Stripe 82 legacy coadded imaging for SN host galaxies. Galaxies selected using our SN sample are shown in black; in blue, the same but only including galaxies known from the IAC Stripe 82 legacy catalogue; in green, the same but only including galaxies known from the SDSS Stripe 82 galaxy catalogue.}
    \label{fig:depth}
\end{figure}

Figure \ref{fig:depth} shows the galaxy magnitude distributions of the 3 SN-matched catalogues presented in this work, giving a clear comparison of their depth. Recall that the SDSS sample was selected to include only \textit{r} < 22.0 (extinction corrected) galaxies. Thus, the best direct comparison of \textit{r}-band magnitude depth is between the IAC data set and that of the present work. The (90th,99th) percentiles of galaxy counts come at \textit{r}-band magnitudes of (22.0,23.8) for the IAC sample and (22.8,25.3) for our bespoke sample, respectively.

\begin{figure}
	\centerline{\includegraphics[width=1.2\columnwidth]{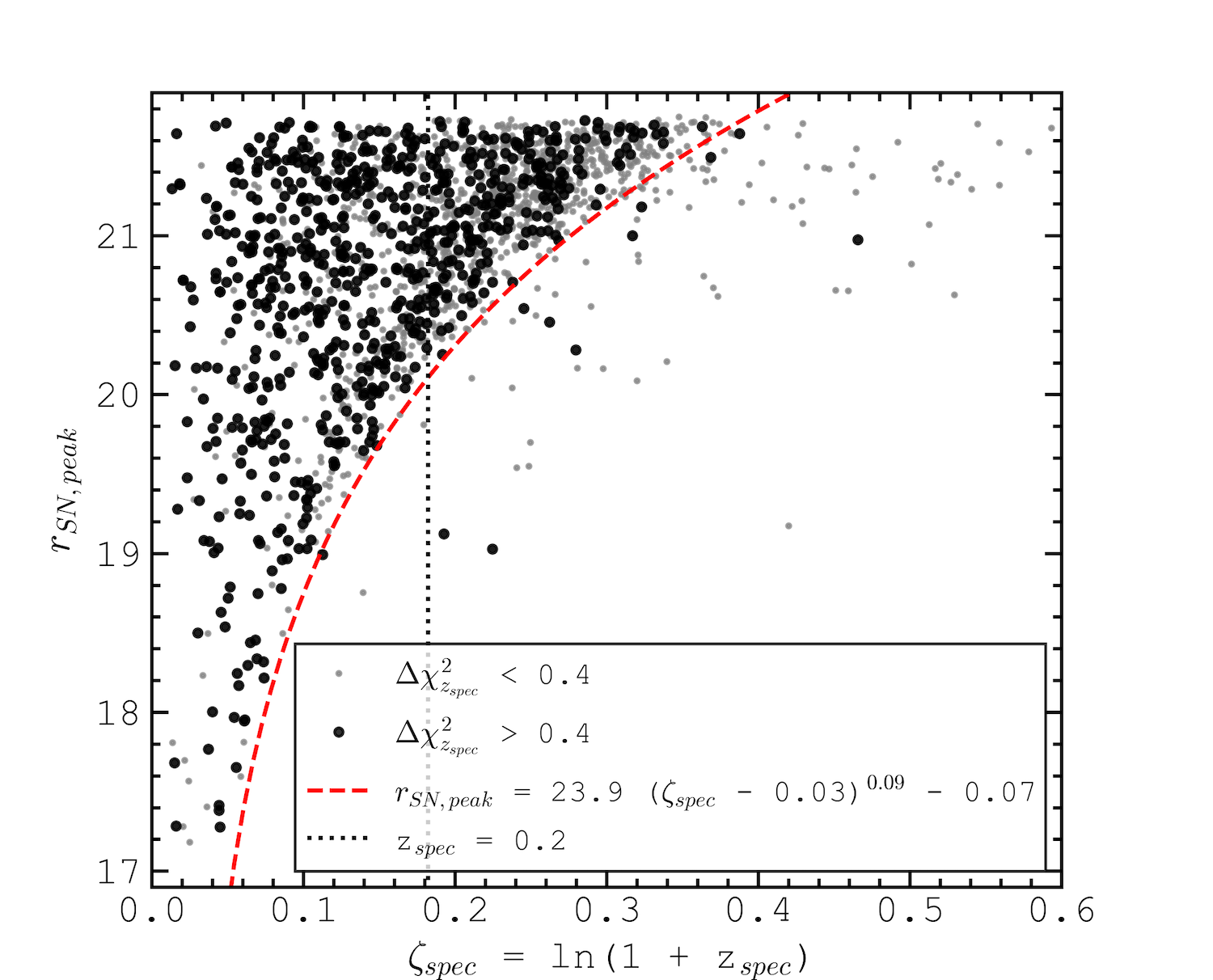}}
    \caption{SN host galaxy spectroscopic redshift, expressed as $\zeta_{spec}$ = ln(1+z$_{spec}$), versus supernova peak \textit{r}-band apparent magnitude ($r_{SN,peak}$), corrected for Galactic extinction. Black circles represent host galaxies with $\Delta \chi_{z_{spec}}^2$ > 0.4, representing more confident redshift estimates. Grey circles represent host galaxies with spectroscopic redshifts not satisfying this criterion. The dashed red line is the inferred line of maximum redshift as a function of $r_{SN,peak}$, used to assess the validity of host galaxy identification.}
    \label{fig:SNmagvsz} 
\end{figure}

We are able to test the success of our galaxy matching by using a relationship between Galactic extinction-corrected SN peak apparent magnitude ($r_{SN,peak}$) and redshift. Figure \ref{fig:SNmagvsz} shows this relationship for those SNe matched to a galaxy for which spectroscopic redshift (z$_{spec}$) is known. Redshift is plotted as $\zeta = \ln(1+z)$ \citep{Baldry18}. For type Ia SNe, there is an approximate maximum SN redshift as a function of SN peak apparent magnitude. CCSNe do not exhibit a relationship that is sufficiently sharp to set a strict maximum redshift, but still follow the same underlying distribution in $r_{SN,peak}$--$\zeta$ space. We can hence test if a galaxy match is reasonable by verifying that the galaxy spectroscopic redshift lies below the maximum redshift expected for that supernova.

For all host galaxies with redshifts exceeding their predicted maximum, \new{given by the red demarcation line in} Figure \ref{fig:SNmagvsz}), we perform further inspection of the coadded imaging, to check for incorrect identification of SN hosts. However, in no case do we find a more sensible host galaxy according to this extra criterion.

Not only can Figure \ref{fig:SNmagvsz} be used to test galaxy matches, it can also be used as a further check that all of our transient objects are indeed SNe. \new{The demarcation line} implies that the maximum redshift for the faintest SNe in our bespoke sample, at $r_{SN,peak} = 21.8$, is z = 0.48. This is hence the maximum trustworthy redshift for SNe in our sample. \citet{BOL12} classify objects as QSOs from their spectra. Cross-matching classifications with our host galaxies, we find a sudden rise in QSO classification at z = 0.48, with $\gtrsim80\%$ of z > 0.48 galaxies classified as QSOs. We thus remove from the SN sample 83 transients assigned to a galaxy with z$_{spec}$ > 0.48, citing the fact that the transient is AGN in nature and not a SN, leaving a \new{final} sample of 2456 host galaxies for CCSNe or type Ia SNe.

\begin{figure}
	\centerline{\includegraphics[width=1.2\columnwidth]{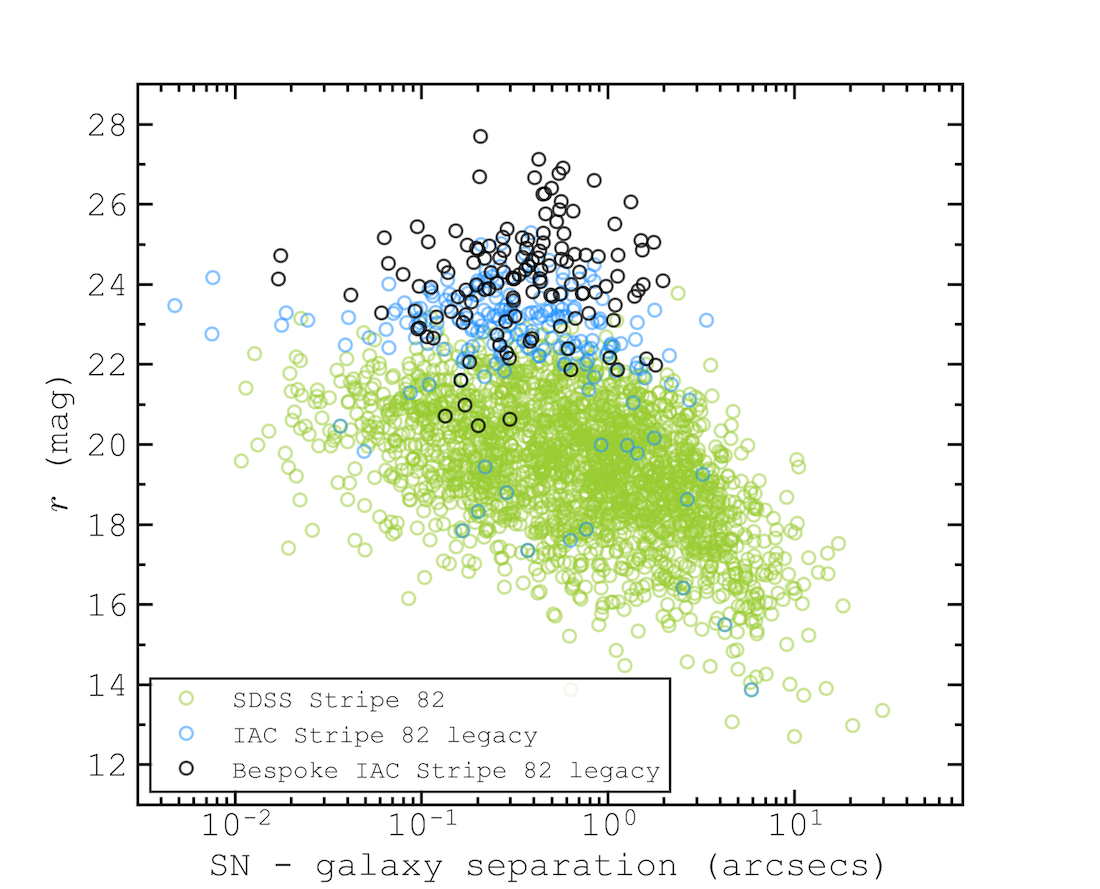}}
    \caption{SN--galaxy separation in arcseconds versus galaxy Kron \textit{r}-band magnitude calculated from our bespoke search of IAC Stripe 82 legacy coadded imaging for SN host galaxies. In black is shown galaxies found using our bespoke SN-host search only; in blue, those found by the IAC Stripe 82 legacy survey; and in green, those found by the SDSS Stripe 82 survey.}
    \label{fig:magsep}
\end{figure}

Figure \ref{fig:magsep} shows SN-Galaxy separation in arcseconds plotted against Kron galaxy magnitude. The same overall distribution is followed for each of the 3 galaxy subsamples: SDSS galaxies, IAC galaxies, and our newly identified LSBGs. SN-galaxy separation increases towards brighter magnitudes due to the correlation of galaxy radius with galaxy magnitude. Normalising separation by galaxy radius, no correlation is found between magnitude and separation. This helps confirm that the SN-galaxy separations found for our LSBGs are reasonable.

To summarise, we have successfully located the host galaxy for each of the supernovae in the sample, with great care taken to ensure the correct host is chosen. \new{Following several steps taken to remove contamination to the SN sample from AGN and variable stars, a sample of 2456 SN host galaxies is obtained. These galaxies were identified 
from single-epoch SDSS imaging (2046), the standard pipeline of multi-epoch Stripe 82 legacy imaging (262), and our bespoke search for the hosts within the multi-epoch imaging (148).} We can now use the resultant galaxy sample, free of incompleteness, for the remainder of our analysis.

\subsection{Stellar masses from observed colours \& redshifts}

Following the determination of the sample we now require galaxy stellar masses, in order to calculate CCSN-rate densities, star-formation rate densities, and star-forming galaxy number densities as a function of galaxy stellar mass. It is often useful to define an approximate stellar mass that can be obtained from absolute magnitudes and colours \citep{BEL03}. This works reasonably well because of the correlation between mass-to-light ratio and rest-frame color \citep{TAY11}. Following \citet{BRY15}, one can also effectively fold in the $k$-correction, and estimate a stellar mass from distances and observed magnitudes as follows: 
\begin{equation}\label{eq:mass}
  \log \mass = -0.4 i + 0.4 {\cal D} + f(\mathrm{z}) + g(\mathrm{z}) (g-i)_{\rm obs}
\end{equation}
where $\mathcal{D}$ is the distance modulus, and $f$ and $g$ are two functions to be determined. 

To calibrate the observed-magnitude-$\mass$ relation, we used the GAMA stellar masses that were determined by spectral-energy-distribution fitting \citep{TAY11,BAL18}. The data were binned in redshift over the range $0.002 < $ z $ < 0.35$, in 18 regular intervals of redshift. The values for $f$ and $g$ were determined for each bin, and finally $f($z$)$ and $g($z$)$ were fitted by polynomials. The coefficients obtained were (1.104,  --1.687,  9.193, --15.15) for $f$ and (0.8237, 0.5908, --12.84,  26.40) for $g$, with the constant terms first. Note because these are cubic functions, they cannot be reliable extrapolated to $z > 0.35$.

Clearly this prescription requires a galaxy redshift in order to calculate stellar mass. We use the most reliable redshift available for our galaxies, as explained below.

\subsubsection{Spectroscopic Redshifts}\label{sec:zspec}

Where available, the most reliable redshifts for our galaxies are spectroscopic. The order of preference for the spectroscopic redshift (z$_{spec}$) used is as follows: 

\begin{enumerate}
    \item Galaxy redshifts obtained by either the SDSS-II Legacy \citep{YOR00}, SDSS-II Southern \citep{BAL05}, or SDSS-III BOSS/SDSS-IV eBOSS surveys \citep{DAW13,DAW16}, and derived via a $\chi^2$ minimisation method described in \citet{BOL12} ($\sim60\%$ of the galaxy sample). 
    \item In the absence of galaxy spectra, we use supernova spectroscopic redshifts, from the various sources outlined in S18 (a further $\sim8\%$ of the sample).
    \item With neither available, in cases where we are confident that the host galaxy which was missed by SDSS is tidally interacting with a galaxy possessing a spectroscopic redshift, we use that spectroscopic redshift ($\sim1\%$ of the sample).
\end{enumerate}

Approximately 70\% of our galaxy sample have some form of spectroscopic redshift. For the remaining galaxies we turn to photometric redshift estimations, described below.

\subsubsection{Photometric Redshifts using zMedIC}\label{sec:zMedIC}

Approximately 65\% of the photometry-only galaxies have SDSS KF-method (kd-tree nearest-neighbour fitted) photometric redshift (z$_{phot}$) estimates \citep{CSA07,BEC16}. However, for galaxies only found in the IAC catalogue or from our bespoke \texttt{SExtractor} implementations, no galaxy redshift is available. A supernova photometric redshift is only known for $\sim$50\% of these galaxies. We set out to calculate galaxy photometric redshifts for the remainder of the sample using a new empirical method. We ultimately find this method to be the most reliable photometric redshift estimator of the three, and hence use our own photometric redshifts for all of the  photometry-only galaxies.

For our photometric redshift determination, we use the IAC Stripe 82 legacy imaging \citep{FT16} along with all available spectroscopic redshifts in Stripe 82. 2.5" circular-aperture-derived \textit{ugriz} colours (used to maximise signal to noise) and their errors are utilised for the method, corrected for Galactic extinction as a function of position, using the extinction maps of \citet{SFD98}.

We use a training set to calculate an empirical relationship between galaxy colours and redshift. The training set consists of a random 50\% of the galaxies for which all of the following conditions are met:
\begin{enumerate}
    \item SDSS/BOSS spectroscopic galaxy redshift is known.
    \item The difference between the $\chi^2$ values of the most likely and second most likely redshift, i.e. $\Delta \chi^2$, is > 0.4. The most likely redshifts were determined from best-fit redshift templates \citep{DAW13}.
    \item Galaxy 2.5" aperture magnitudes have S/N > 10 in each of the optical Sloan filters, corresponding to colour errors of $\lesssim$0.15 magnitudes.
\end{enumerate}

Conditions (ii) to (iii) ensure the spectroscopic redshifts and colours we train our colour-z relation on are reliable. The resulting training and validation sets each consist of $\sim$22000 galaxies.

Equation \ref{eq:zMedIC} (below) is used to relate galaxy colours to redshift, giving each colour a coefficient. The optimal coefficients are those which yield an output z$_{phot}$ which resembles the known z$_{spec}$ for our training sample galaxy. The coefficients $k_{1}$ and $k_{2}$ in Equation \ref{eq:zMedIC} are used as scaling values, to normalise and linearise the relationship between colours and z$_{phot}$. We name this code zMedIC (\textbf{z} \textbf{Me}asure\textbf{d} via \textbf{I}teration of \textbf{C}oefficients).

\begin{equation}\label{eq:zMedIC}
\zeta_{phot} = \left (c_1 (u-g) + c_2 (g-r) + c_3 (r-i) + c_4 (i-z) + k_1 \right )^{k_2}
\end{equation}

Optimal coefficients are found by producing random sets of numbers to iteratively approach the set which minimises the $\zeta_{spec}$ root-mean-square (rms) deviation, given by Equation \ref{eq:sigma}. 

\begin{equation}\label{eq:sigma}
\sigma = \sqrt{\frac{\sum_{i=1}^{n}w_i(\zeta_{phot}-\zeta_{spec})^2}{\sum_{i=1}^{n}w_i}}
\end{equation}
 
 The weights $w_i$ are chosen to give larger weight to confident z$_{spec}$ measurements and more precise colours, and to even out weighting across redshift space. To ensure the latter, we divide the data into $\zeta$ bins of width 0.05, normalising weights with a value, $\mathcal{F}_{bin}$, different for each bin, such that the sum of weights is the same in each bin. Thus, the final form of the weights $w_i$ is $ 1/(\chi_{z_{spec}}^2 \mathcal{F}_{bin} \Delta (i-z))$. Due to the observed upper limit to redshift as a function of SN apparent magnitude, given by \new{the red demarcation line} in Figure \ref{fig:SNmagvsz}, we add a constraint such that the set of coefficients do not result in more than 10\% of the sample lying above this redshift limit.

\begin{figure}
	\centerline{\includegraphics[width=1.2\columnwidth]{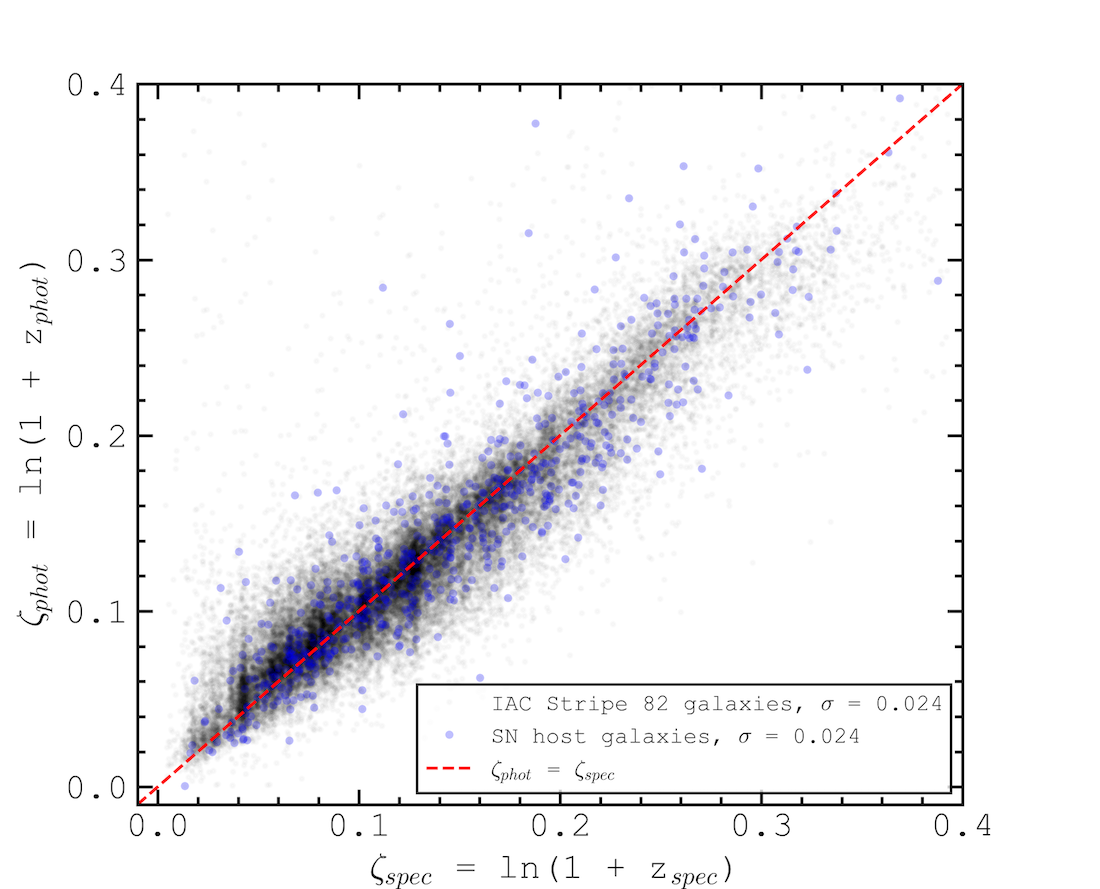}}
    \caption{Spectroscopic versus photometric redshift estimates from zMedIC, derived using 2.5" circular aperture-derived optical galaxy colours. Redshifts are represented as $\zeta$ = ln(1+z). The validation set of $\sim$22000 galaxies from the IAC Stripe 82 Legacy catalogue is shown as black circles. The validation set of $\sim$400 SN host galaxies found from a bespoke search of the same IAC coadded imaging is shown as cyan circles. All redshifts are calculated using the zMedIC coefficients found from the training set of $\sim$22000 galaxies. Weighted root-mean-square (rms) deviations ($\sigma$) in $\zeta$ are indicated.}
    \label{fig:zMedIC}
\end{figure}

Figure \ref{fig:zMedIC} shows $\zeta_{spec}$ vs $\zeta_{phot}$ for each \textit{r}-band magnitude bin, and the corresponding values of $\sigma$ for our validation set. The best-fit coefficients of Equation \ref{eq:zMedIC} are shown in Table \ref{tab:zMedIC}. Note that this colour-z relation is trained on both star-forming and quiescent galaxies. Removing quiescent galaxies from the training set to better reflect our SN hosts, we find no significant change to the optimal coefficients. Furthermore, training instead on only the SN sample of host galaxies for which the above training set selection criteria are met ($\sim$400 galaxies), coefficients are similarly unaffected. We also test the effect of binning galaxies by $r$-band apparent magnitude, to check for potential improvements to $\sigma$. No benefit to $\sigma$ is found when dividing the large training sample of $\sim$22000 galaxies into magnitude bins, but for the much smaller sample of $\sim$400 CCSN-host galaxies, binning by magnitude is found to reduce $\sigma$ for the $r \leq 19.5$ galaxies by $\sim$20\%.

The redshift parameterisation of Equation \ref{eq:zMedIC} can be modified to deal with non-detections. The optimal coefficients are calculated with every possible combination of colours. For instance, if a non-detection is found for a galaxy in the $g$-band only, we are able to use the optimal coefficients associated with the colours ($u-r$), ($r-i$) and ($i-z$), in order to infer photometric redshift. The most common filter with non-detections for our host galaxies is the $u$-band. Removing $(u-g)$ colour increases $\sigma$ by $\sim$20\% and $\sim$2\% for $r \leq 19.5$ and $r > 19.5$.

\begin{table}
\centering
\caption{Best-fit values for the coefficients in Equation \ref{eq:zMedIC}, as inferred from a training set of $\sim$22000 galaxies, with coefficients calculated separately for 3 bins of \textit{r}-band galaxy magnitude, and for the entire training sample (labelled \textquotesingle All\textquotesingle).}
\label{tab:zMedIC}
\resizebox{\columnwidth}{!}{
\begin{tabular}{c|cccccc|c}
 & $c_1$ & $c_2$ & $c_3$ & $c_4$ & $k_1$ & $k_2$ \\ \hline
All & --0.21 & 0.32 & --0.11 & --0.03 & 0.18 & 0.90 \\
$r$ $\leq$ 17.5 & --0.17 & 0.55 & 0.13 & --0.38 & 0.18 & 1.97 \\
17.5 < $r$ $\leq$ 19.5 & --0.15 & 0.35 & --0.22 & --0.12 & 0.13 & 0.74 \\
$r$ > 19.5 & --0.19 & 0.38 & --0.07 & 0.06 & 0.12 & 0.87
\end{tabular}
}
\end{table}

The method works over the redshift range of 0 < z < 0.4. Beyond this upper limit, there are too few reliable spectroscopic redshifts in our sample for an assessment, and the chosen relationship for all galaxies may be expected to break down due to evolutionary effects and more complex k-corrections.

To emphasise, we find that Equation \ref{eq:zMedIC} yields the most accurate redshift estimates when using galaxy colours alone. The coefficients used are specific to 2.5" circular aperture colours. The rms deviation values between photometric and spectroscopic redshifts for this method are comparable to those using the SDSS KF-method \citep{CSA07,BEC16}, and are substantially better than rms deviations using a SN light-curve method \citep{GUY07}.

With redshifts estimated, we now remove likely type Ia supernovae from our sample to leave only likely CCSNe. In Section \ref{sec:CCSNRD} we compare the effects of assumptions for the nature of the unknown-type SNe. It is likely that the unknown-type fraction of the sample consists of both CCSNe and type Ia SNe, and it is discussed how we attempt to circumvent this problem.

\subsubsection{Monte Carlo Assessment of Uncertainties} \label{sec:MC}
\indent The CCSN-rate densities, star-formation rate densities and hence galaxy stellar mass functions we will derive are sensitive to redshift estimates in 2 ways: \new{i) galaxy stellar masses are estimated using redshifts (see Equation \ref{eq:mass}); ii) incompleteness of the sample is a function of redshift. Redshifts are required to volume limit the sample}. As small numbers of log($\mass$/M$_{\odot}$) $\lesssim$ 8.0 galaxies can significantly change the form of the SFRD, and hence the GSMF, we must take care when utilising photometric redshifts.

\begin{figure}
	\centerline{\includegraphics[width=1.2\columnwidth]{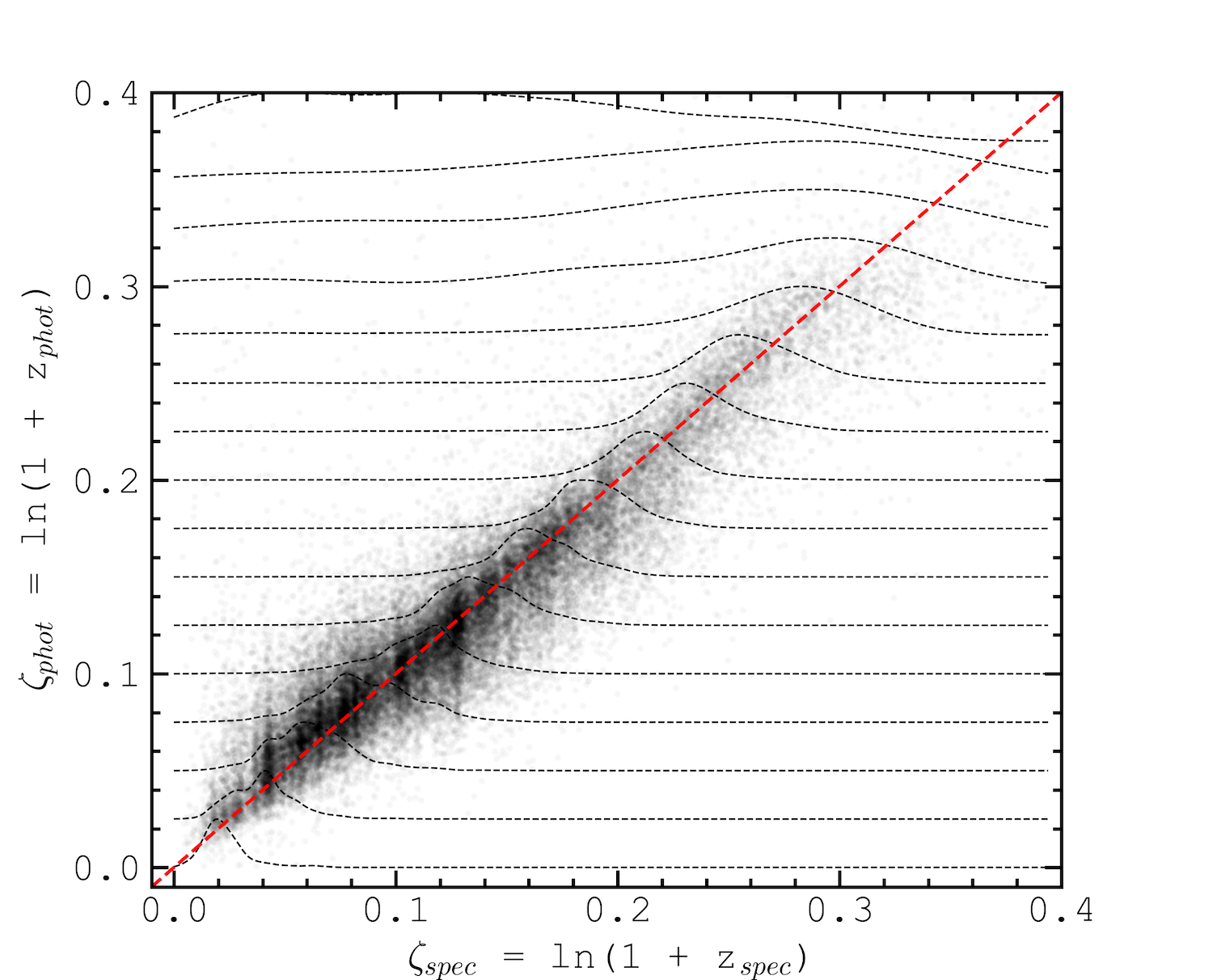}}
    \caption{As in Figure \ref{fig:zMedIC}, but showing only the validation set of approximately $\sim$22000 IAC Stripe 82 legacy galaxies. Probability density functions (PDFs) of spectroscopic redshift as a function of photometric redshift are superimposed; these were used to model photometric redshift uncertainties. PDFs are calculated from a multi-Gaussian kernel-density-estimation, in $\zeta_{phot}$ bins of width 0.025. Redshift PDFs as a function of photometric redshift estimation are implemented into a 1000-iteration Monte Carlo method to assess galaxy stellar mass uncertainties.}
    \label{fig:PDFs}
\end{figure}

To circumvent this problem, we turn to a Monte Carlo technique. Firstly, to assess the uncertainty in our photometric redshifts we divide the $\zeta_{spec}$ vs $\zeta_{phot}$ space into bins along the $\zeta_{phot}$ axis. We then make a histogram of counts along the $\zeta_{spec}$ axis for each $\zeta_{phot}$ bin, smoothed via a multi-Gaussian kernel-density estimation (kde) \citep{PAR62}. As such, we have the $\zeta_{spec}$ distribution as a function of $\zeta_{phot}$. We can use each kde as a probability density function (PDF) for a particular \new{$\zeta_{spec}$} distribution as a function of our $\zeta_{phot}$ input. These distributions are shown superimposed onto the combined $\zeta_{spec}$ vs $\zeta_{phot}$ distribution in Figure \ref{fig:PDFs}. \new{$\zeta_{phot}$ is found to be systematically greater than $\zeta_{spec}$ at the highest values {($\zeta_{phot}$ $\gtrsim$ 0.3)}. The PDF is designed to statistically account for this systematic when implemented into the Monte Carlo technique.}

We then run a 1000-iteration Monte Carlo code where each photometric-galaxy\textquotesingle s redshift is replaced by a value drawn from the probability density functions of redshift shown in Figure \ref{fig:PDFs}. For spectroscopic galaxies, the spectroscopic redshift is used, and varied for each Monte Carlo iteration according to its error. For each iteration we convert a galaxy\textquotesingle s z-estimate to a luminosity distance assuming an $h = 0.7$, $\Omega_m = 0.3$, $\Omega_{\Lambda} = 0.7$, flat cosmology. \textit{g} and \textit{i} 2.5" circular-aperture-derived magnitudes, as well as elliptical Kron-aperture-derived $i$-band magnitude, are varied with each iteration in accordance with their uncertainties. This allows an estimate of the galaxy\textquotesingle s stellar mass for each iteration using Equation \ref{eq:mass}. 

To volume limit the sample to obtain a SFRD, a z < 0.2 cut is made following each Monte Carlo iteration. This cut is chosen to limit the effects of galaxy evolution at higher redshifts, to match to the redshift cut of G10, for the most direct comparison of SFRDs, and to limit the effects of extinction on SN counts. \new{At higher redshifts, more SNe are near the detection limit of the survey, where a small amount of extinction can make the SNe undetectable. Limiting the number of SN detections sensitive to extinction decreases our results' reliance on the extinction model}. Using this cut, the number density of CCSNe as a function of host galaxy mass is estimated, with galaxy stellar mass bins 0.2 dex in width. We then take the mean of bin counts and the standard deviation of bin counts over the 1000 iterations, for each bin.

\section{Results \& Discussion}\label{sec:ResultsandDiscussion}

\subsection{Observed CCSN-Rate Densities}\label{sec:CCSNRD}

Using Equation \ref{eq:CCSNobs}, we can convert number statistics of CCSNe as a function of host galaxy stellar mass into \new{volumetric} CCSN-rate densities, given effective SN rest frame survey time $\tau$ and survey volume $V$. CCSN-rate densities are corrected for cosmological time dilation effects on survey time period as a function of redshift.

\begin{figure}
	\centerline{\includegraphics[width=1.2\columnwidth]{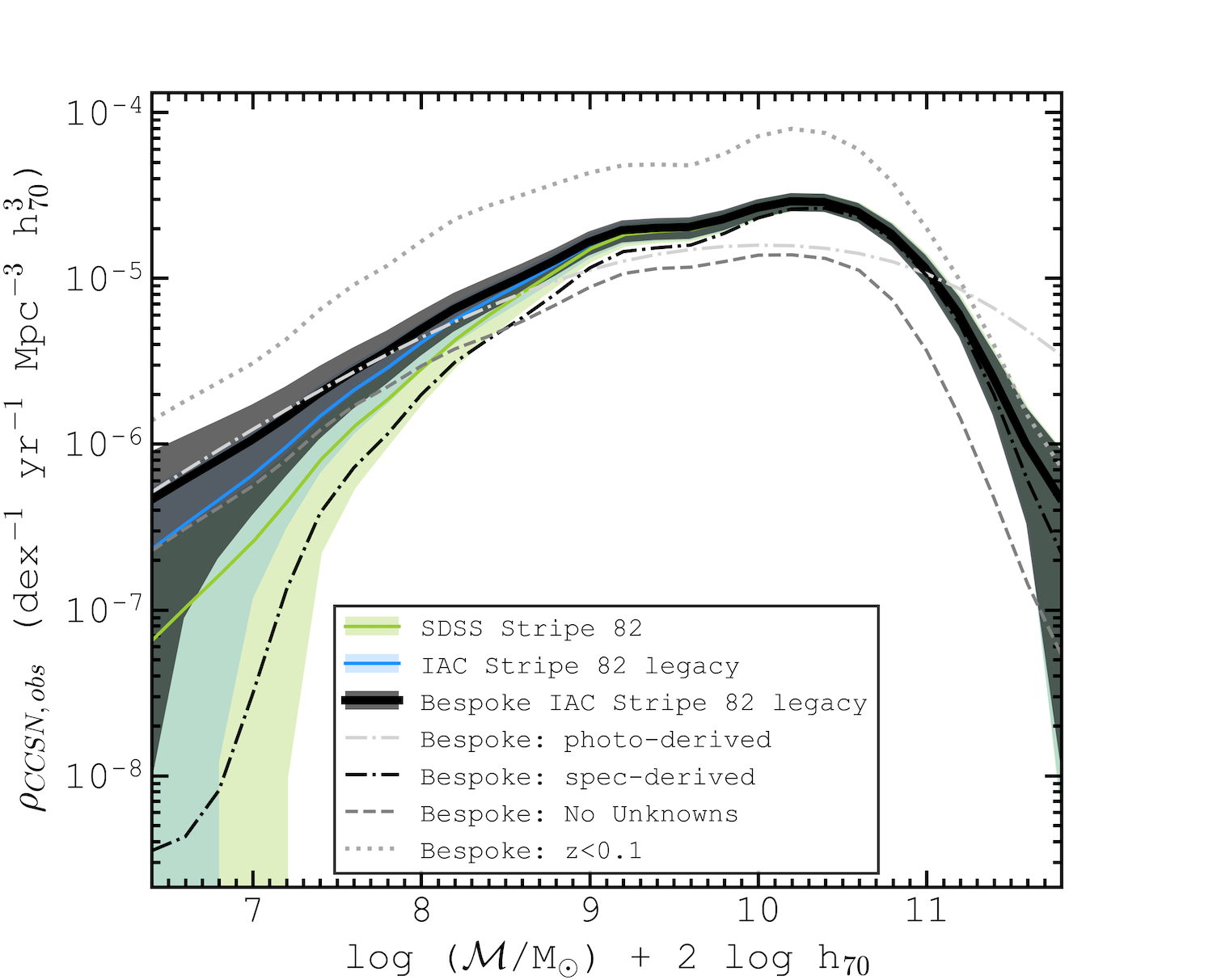}} 
    \caption{\new{Volumetric} CCSN-rate densities (z $<$ 0.2) as a function of host galaxy stellar mass. The black solid line shows galaxies derived from a bespoke search for CCSN host galaxies in IAC Stripe 82 Legacy imaging. The blue line shows CCSN host galaxies known from the IAC Stripe 82 legacy galaxy catalogue. The green line shows CCSN host galaxies known from the SDSS Stripe 82 galaxy catalogue. Shaded regions indicate the 1$\sigma$ of standard deviation from a Monte Carlo method and Poisson errors. The dot-dashed grey line shows the bespoke sample but with all galaxy masses derived using photometric redshifts, while the black dot-dashed line shows all galaxy masses derived using spectroscopic redshifts (with galaxies omitted where spectroscopic redshifts are unavailable). The grey dashed line shows the bespoke sample but with all unknown-type SNe removed. The grey dotted line shows the bespoke sample but with a redshift cut of z $<$ 0.1.}
    \label{fig:rho_CCSN_uncorr}
\end{figure}

Figure \ref{fig:rho_CCSN_uncorr} shows \new{volumetric} z < 0.2 CCSN number densities as a function of host galaxy mass, derived from our Monte Carlo technique. Based on the redshift cut, sky coverage and the effective span of the Stripe 82 SN survey [$\tau \sim 270 / (1+\overline{z})$ days], $\sim 10^{-7}$ CCSN yr$^{-1}$Mpc$^{-3}$h$_{70}^3$ corresponds to 1 observed CCSN. As a result, we do not assess densities below log($\mass$/M$_{\odot}$) = 6.4, as below this mass the mean number counts per bin descend below 1 for the full sample of galaxies found from a bespoke search for SN hosts in IAC Stripe 82 Legacy imaging. For this sample we find CCSN number densities to decrease as a power-law for log($\mass$/M$_{\odot}$) $\lesssim$ 9.0. To show the effects of selection bias we also calculate CCSN number densities using only those CCSNe assigned to hosts in the SDSS and IAC galaxy catalogues. As expected, consistency is found at higher masses, whilst a deviation in CCSN counts is found at lower host galaxy masses (log($\mass$/M$_{\odot}$) < 9.0) due to decreased sample completeness. A double peak in number density is observed, consisting of a primary peak at log($\mass$/M$_{\odot}$) $\sim$ 10.8 and a secondary peak at log($\mass$/M$_{\odot}$) $\sim$ 9.4. Using the alternative stellar mass prescription of G10 in our Monte Carlo method, a modification of the prescription introduced by \citet{KAU03}, we find this double-peak to persist.

Error bars incorporate the uncertainties in redshift and in the stellar mass parameterisation, as well as the uncertainty in the nature of the unknown-type transients. We have built on the work of S18 to deduce that these objects are very likely to be supernovae. However, each of these objects could be CC or type Ia SNe. We use volume-limited SN number statistics, calculated by R14, to derive a ratio of CCSNe to type Ia SNe. This gives a predicted percentage of unknown-type SNe that are type Ia, and for each Monte Carlo iteration, this percentage of unknown-type SNe, selected at random, are removed from the sample.

Figure \ref{fig:rho_CCSN_uncorr} shows the effect of removing this fraction of unknown-type SNe from the sample. Comparing with the full sample, no strong correlation is found .between the percentage of SNe that are unknown-type and galaxy stellar mass. As such, a removal of a percentage of unknown-type SNe to attempt to remove type Ia's effectively corrects number densities by a constant amount, rather than modifying the CCSN-rate density \new{distribution} with mass. Changing the percentage of unknown-type SNe removed does not affect the presence of the double-peak in CCSN-rate density. 

Also shown is the sub-sample of galaxies for which spectroscopic redshift is known. Lower-mass galaxies are less likely to have been selected for spectroscopic analysis. The low-mass limit of the CCSN-rate density is therefore dominated by galaxies with photometric redshifts. 

To test the performance of zMedIC in producing reliable redshift estimates, we observe the effect of calculating all galaxy masses using our photometric redshifts. We see in Figure \ref{fig:rho_CCSN_uncorr}, that zMedIC-derived redshifts, depicted by the grey dot-dashed series, are able to reproduce the fundamental shape of the CCSN number densities as a function of mass, but that fine features such as the double-peak are not reproduced.

Also plotted in Figure \ref{fig:rho_CCSN_uncorr} is the bespoke sample's CCSN-rate densities using a redshift cut of z < 0.1. CCSN-rate densities are increased when using this cut, by a factor of $\sim$3. This is because we do not yet have a truly volume limited sample, due to the supernova magnitude cut. This draws our attention to the need for corrections for SN-detection efficiency, $\epsilon$, as discussed in Section \ref{sec:Corrections}.

\begin{figure}
	\centerline{\includegraphics[width=1.2\columnwidth]{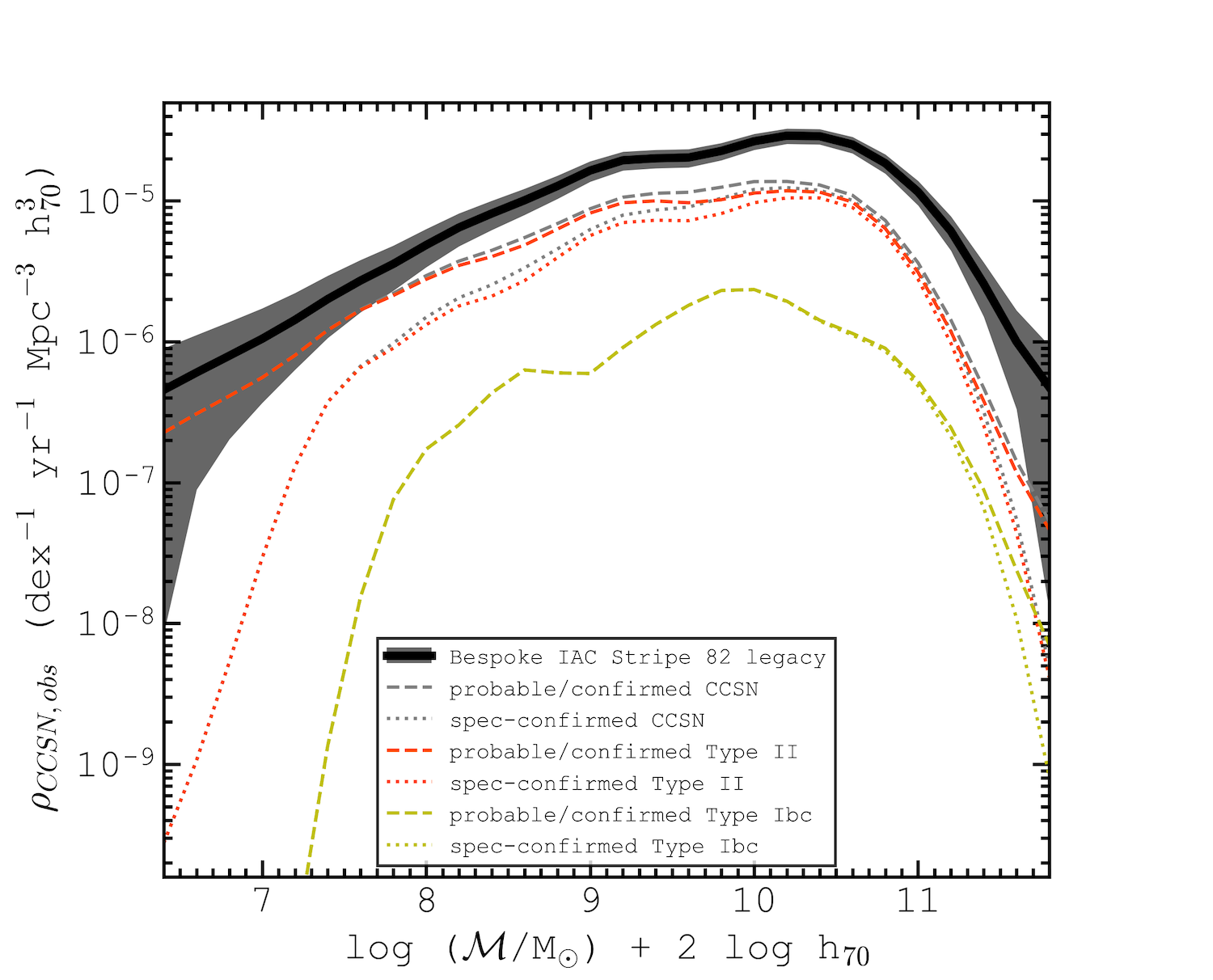}}
    \caption{As Figure \ref{fig:rho_CCSN_uncorr}, but also showing the full CCSN host sample subdivided into probable/confirmed CCSNe (grey dashed) and confirmed CCSNe only (grey dotted), where probable indicates those classified via photometry only, and where confirmed indicates those classified via spectroscopy. These series are subdivided further by SN type: probable/confirmed and confirmed type II SN (red dashed, red dotted) and probable/confirmed and confirmed type Ib/c SN (yellow dashed, yellow dotted).}
    \label{fig:rho_CCSN_types}
\end{figure}

To test if the double-peak in observed CCSN-rate density as a function of galaxy mass arises due to a particular SN type, we plot Figure \ref{fig:rho_CCSN_types}, separating the CCSN sample into type II and type Ib/c SN. Those with spectroscopically confirmed SN types are also plotted in isolation. For all SN types and for probable and confirmed SNe-types, the double-peak remains, with or without a Monte Carlo variation of masses.

\subsection{Corrected CCSN-Rate Densities} \label{sec:Corrections}

In Section \ref{sec:SNmagcut} we found that it is unlikely that any $r_{SN,peak}$ < 21.8 SNe are missed by the instrumentation described in S18 over the observing seasons, and that by locating the host galaxy for each of these SNe, surface brightness/mass biases are significantly reduced. However, the magnitude $r_{SN,peak}$, which controls whether a SN is contained in our sample, may be a function of redshift, the type of core-collapse supernova that we are observing, and, most importantly for this analysis, host galaxy stellar mass.

\begin{enumerate}
\item \textbf{In the case of galaxy mass}, higher mass galaxies may contain more dust than lower mass galaxies. G10 uses the Balmer decrement to estimate the dependency of A$_{H\alpha}$ on galaxy stellar mass. We test the assumption that rest frame \textit{r}-band extinction for a SN line-of-sight in its host galaxy follows the same extinction-mass relation as the H$\alpha$ line. Alternatively, extinction in CCSN environments within these galaxies may not be so strongly dependent on the extinction inferred from the Balmer decrement. In our sample, mean CCSN colours at peak epoch do not show any notable correlation with host mass. For type Ia SNe, which are known to peak at approximately the same (B--V) colour \citep{NUG02}, there is no correlation between colour and host mass (although environments may differ for type Ia and CCSNe).

\item \textbf{In the case of redshift}, CCSNe will be fainter with distance due to inverse square dimming to flux. They will also generally experience additional dimming to \textit{r}-band magnitude with redshift, in our low redshift regime, due to the shape of their spectra. k-corrections are therefore necessary to represent the higher redshift portion of the sample correctly. CCSN k-corrections are estimated using a type IIP template spectrum from \citet{GPN99} at peak magnitude for all SNe in our sample, and following equation~1 of \citet{KIM96}. 
\end{enumerate}
To investigate these effects, we consider how SN detection efficiency depends on host galaxy mass and redshift. Using absolute \textit{r}-band magnitude distributions of SNe as a function of SN type, we can estimate the probability that each SN is brighter than $r_{SN,peak}$ = 21.8, given its redshift and extinction. These probabilities lead to detection efficiencies, $\epsilon$. Each CCSN contributes 1/$\epsilon$ counts to the number density within its galaxy stellar mass bin, leading to corrected CCSN-rate densities. Detection efficiencies are calculated for each Monte Carlo iteration to account for the uncertainties in redshift, mass and SN type that efficiencies depend on.

In order to estimate detection efficiency we first require an assumption about SN absolute magnitudes. The distributions used are derived from the volume-limited distributions calculated by R14, who observe approximately Gaussian distributions for each SN type. These SN types are finer classifications than made in S18: a type II SN as classified by S18 could be any one of IIP, IIL, IIn or IIb in R14. Therefore if each of the 4 sub-types follow different Gaussian distributions in absolute magnitude, we assume an absolute magnitude for all type II SNe which is the sum of these 4 Gaussians, whilst preserving the relative counts for each sub-type. This is done similarly for type Ib and type Ic SNe, which come under \textquotesingle type Ib/c\textquotesingle\; in S18. The $r$-band absolute magnitude distributions used are derived from the $B$-band distributions of R14, converted using the prescription of \citet{JES05} for stars with $R_c$ -- $I_c$ < 1.15. Assuming ($B$--$V$) = 0.0, as found to be typical for type Ia SNe at peak magnitude by \citet{NUG02}, we estimate that $M_r \sim M_B + 0.1$ for our CCSNe. 

If the absolute magnitude distribution of a SN type can be approximated as a Gaussian with a mean $\overline{M}$ (and standard deviation $\sigma$), then for an individual SN of that type, the mean apparent magnitude $\overline{m}$ as a function of redshift and extinction is given by Equation \ref{eq:SNcorrections1}, where $k_{rr}$ is the $r$-band k-correction for the SNe, A$_{r,h}$ is the host galaxy $r$-band extinction and A$_{r,MW}$ is the Galactic $r$-band extinction along the line-of-sight.

\begin{equation}\label{eq:SNcorrections1}
\overline{m} = \overline{M} + 5 \log d_{L}(\mathrm{z}) + k_{rr} + A_{r,h} + A_{r,MW}
\end{equation}

Again, we take the sum of the Gaussian distributions to obtain distributions for type Ib/c and type II SNe. By integrating under the summed Gaussian distributions, the efficiency of detection, $\epsilon$, as a function of redshift and extinction, correcting for the effects of a SN peak magnitude cut at $r_{SN,peak} = 21.8$ is then estimated by Equation \ref{eq:SNcorrections2}, where $N_i$ are the predicted relative numbers of each SN type, used to weight the sum of the $n$ Gaussians.

\begin{equation}\label{eq:SNcorrections2}
\epsilon=\frac{1}{2} - \frac{1}{2\sum_{i}^{n}N_{i}}\sum_{i}^{n}N_{i}\erf \left (\frac{\overline{m_{i}}-21.8}{\sqrt{2}\sigma_{i}}  \right )
\end{equation}

Detection efficiency is clearly a function of SN type. We test the effects of bias in the SN classifications of S18 by varying the ratio of type Ia, Ib/c and II SNe in the unknown-type fraction of the sample. However, it is found that any effects are of second-order importance. Therefore, we simply assume that the unknown-type SNe follow the volume-limited type ratios of R14.

For each Monte Carlo iteration, unknown-type SNe are reassigned a SN type, and are thus either removed from the sample as a type Ia, or are given an absolute magnitude drawn from the distribution associated with either a type Ib/c or type II SN, which enables an estimate of their detection efficiency for each iteration.

\begin{figure}
    \vspace*{-0.00cm}
	\centerline{\includegraphics[width=1.2\columnwidth]{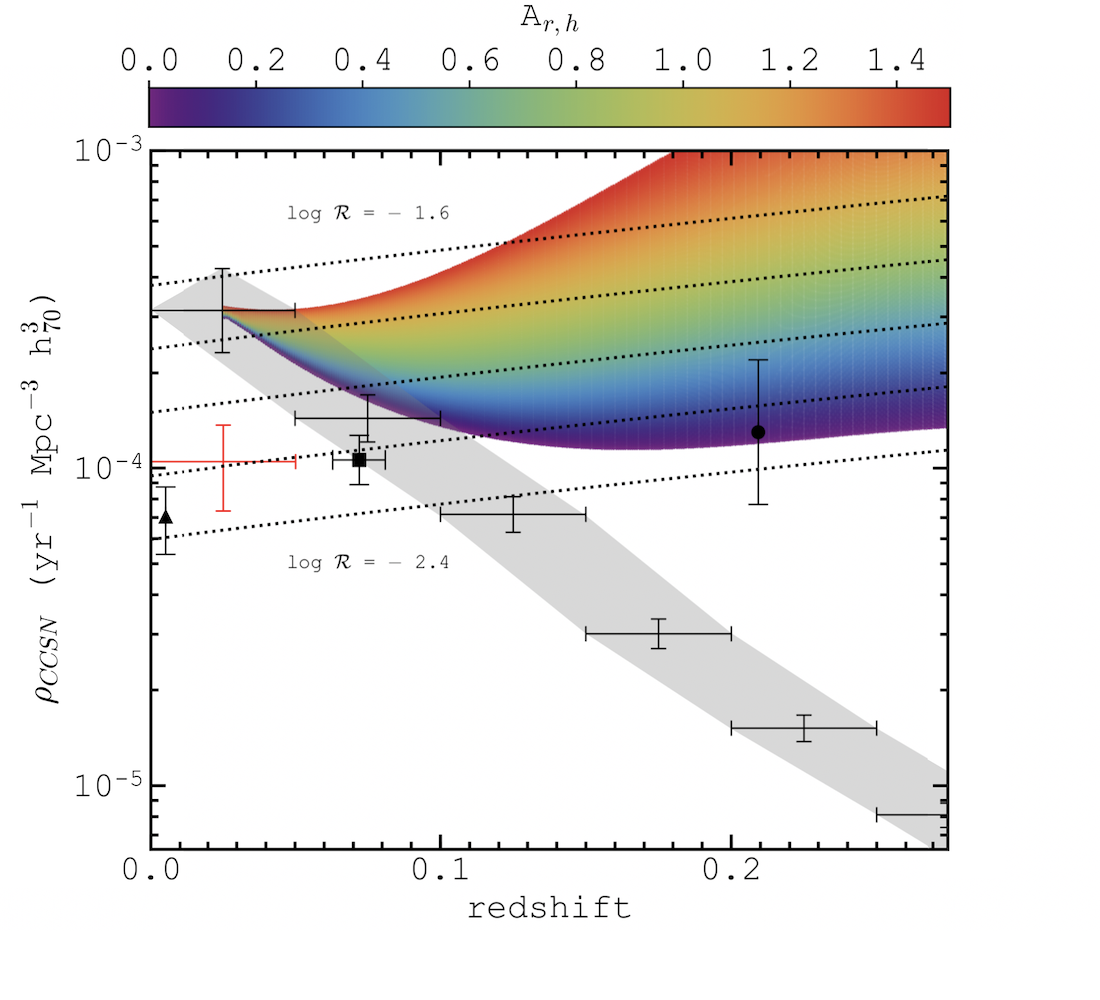}}
    \caption{\new{Volumetric} CCSN-rate density versus redshift, calculated in redshift bins of 0.05, and as a function of SN detection efficiency corrections. The grey shaded region depicts observed CCSN-rate densities from our CCSN sample, uncorrected for SN detection efficiencies, bounded by 1$\sigma$ Poisson+Monte Carlo+Cosmic Variance errors. Colour-shaded density values are corrected for SN detection efficiency. The colour bar represents the different corrected CCSN-rates obtained when assuming different values of CCSN host galaxy extinction, A$_{r,h}$. Dashed lines represent expected CCSN-rate histories derived from the star-formation history of \citet{MD14}, assuming different values of log $\mathcal{R}$ in the range --1.6 to --2.4 (see text). Triangular, square and circular points represent CCSN-rate densities obtained by \citet{LI11}, \citet{TAY14} and \citet{BOT08}, respectively. \new{The red point shows how our observed $z<0.05$ SN counts are reduced when objects inferred to have peak absolute magnitudes $> -15$ are discounted as CCSNe.}}
    \label{fig:rho_CCSN_z}
\end{figure}

Equations \ref{eq:SNcorrections1} and \ref{eq:SNcorrections2} are sensitive to assumptions for host galaxy extinction, A$_{r,h}$. Significant uncertainty exists around the relationship between CCSN extinction and host galaxy mass. Figure \ref{fig:rho_CCSN_z} shows corrected CCSN-rate density vs redshift, for different values of A$_{r,h}$. In comparison to the observed values, clearly there are larger corrections at higher redshift and with higher assumed A$_{r,h}$. 

Also shown in Figure \ref{fig:rho_CCSN_z} is the inferred CCSN-rate density derived from the star-formation history of \citet{MD14}, assuming different values for $\mathcal{R}$, the expected number of stars that explode as CCSNe per unit mass of stars formed. \citet{MD14} assume a Salpeter IMF for the star formation history. Using this IMF with initial masses 0.1 -- 100 M$_{\odot}$ and assuming that all stars with initial masses 8 -- 40 M$_{\odot}$ result in CCSNe, then $\log \mathcal{R} = -2.17$. 
We find that using values of A$_{r,h}$ from $\sim 0.3$ to 0.6 reproduces the evolution of CCSN density with redshift, with $\log \mathcal{R}$ in the range $-2.2$ to $-1.8$. \new{We adopt $\log \mathcal{R} = -1.9$}. 

Note though that the measured $z<0.05$ rate is higher than expected. \new{We find that there is a significant excess of SNe with very faint peak $r$-band absolute magnitudes ($M_{r,sn}>-15$), only found for this redshift bin. R14 predict the fraction of CCSNe with $M_{r,sn}>-15$ to be negligible. One explanation for this excess could be a contamination of the SN sample at these lowest redshifts from outbursts of Luminous Blue Variables, which can exhibit similar light curve properties to Type IIn supernovae \citep[see, e.g.][]{GOO89,VD00}. The red point of Figure \ref{fig:rho_CCSN_z} shows the effect of removing objects inferred to have $M_{r,sn}>-15$ on the $z<0.05$ CCSN counts. This corrected value is in agreement with \citet{LI11} and \citet{TAY14}.}

\begin{figure}
	\centerline{\includegraphics[width=1.2\columnwidth]{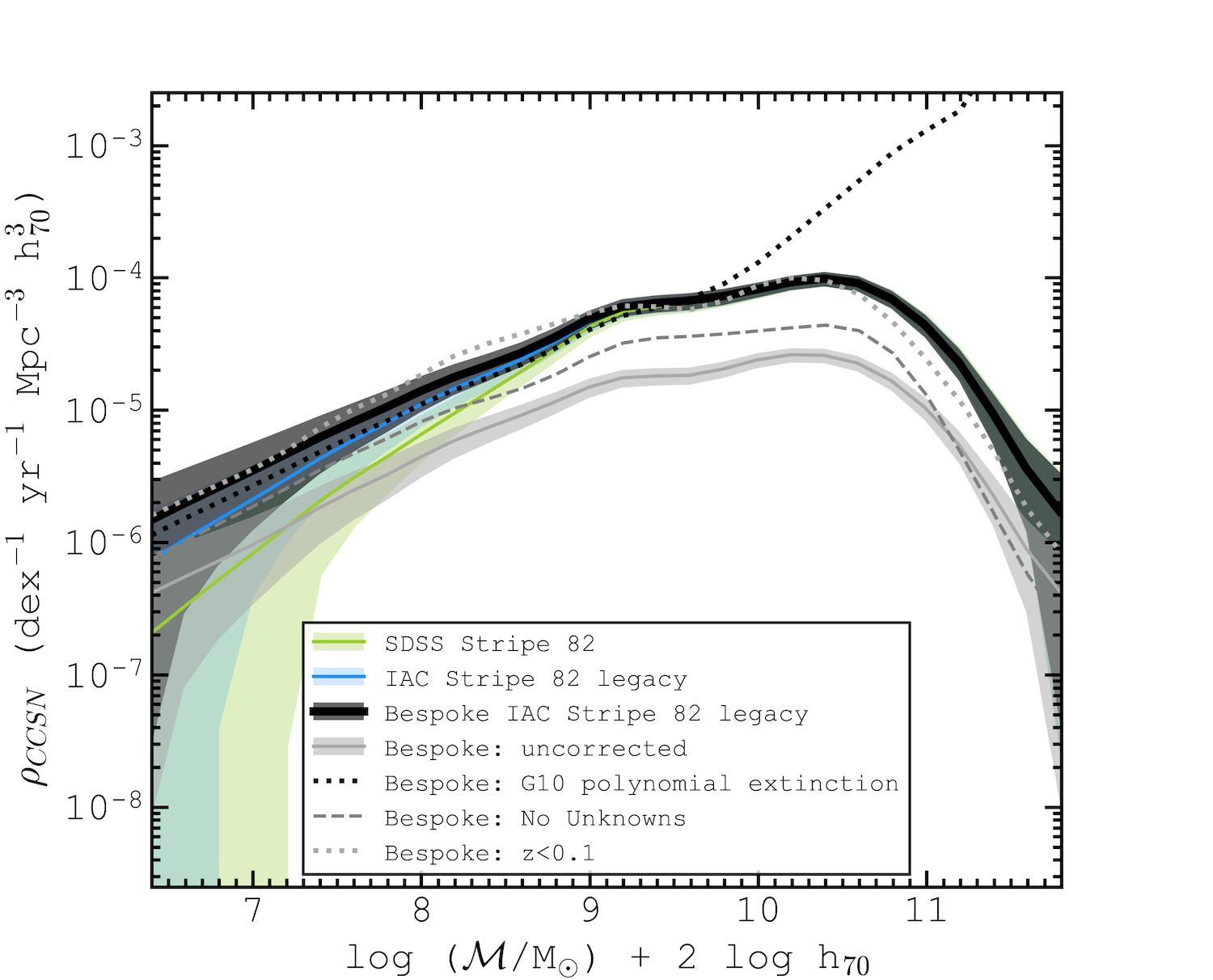}}
    \caption{\new{Volumetric} z $<$ 0.2 CCSN-rate densities as a function of host galaxy stellar mass, corrected for the effects of SN detection efficiencies. All series are as described in Figure \ref{fig:rho_CCSN_uncorr}. All results assume constant host galaxy extinction with mass, A$_{r,h}$=0.5 mag, except the black dotted line which assumes polynomially increasing host galaxy extinction with mass (\new{as in G10}). The grey solid line and shaded region represent CCSN-rate densities uncorrected for SN detection efficiencies and their 1$\sigma$ errors, respectively.}
    \label{fig:rho_CCSN_corr}
\end{figure}

The effects of corrections for SN detection bias on CCSN number densities as a function of galaxy stellar mass are shown in Figure \ref{fig:rho_CCSN_corr} (with A$_{r,h} = 0.5$). CCSN-rate densities (z < 0.2) are now higher than the uncorrected values by a factor of $\sim 2$ at the low-mass limit, and by a factor of $\sim 3$ at log($\mass$/M$_{\odot}$) = 10.6, the position of the primary peak. Even though host galaxy extinction is set constant with galaxy stellar mass, this larger factor at higher masses indicates the effects of a weak SN-type dependence on galaxy stellar mass.

The uncertainties take account of the efficiency corrections in Figure \ref{fig:rho_CCSN_corr}. Even with these uncertainties, corrected CCSN-rate densities \new{using the full sample of CCSN hosts} show a power-law decrease with decreasing stellar mass, down to the low-mass limit of the sample. 
This is not evident from the samples that do not use the bespoke LSBG detections.
The 1$\sigma$ levels indicated by the shaded regions suggest non-zero number densities from single-epoch SDSS imaging only down to log($\mass$/M$_{\odot}$) = 7.2, and down to log($\mass$/M$_{\odot}$) = 6.8 for IAC Stripe 82 coadded imaging. The z < 0.1 densities now approximate the z < 0.2 densities across the mass range, indicating the validity of corrections. 

The black dotted line of Figure \ref{fig:rho_CCSN_corr} shows CCSN-rate densities assuming increasing extinction with mass as in G10. High mass counts appear to be significantly overestimated compared to predictions for the star-formation rate density (see Section \ref{sec:SFRD}). This discrepancy is not surprising because G10 uses the Balmer decrement and we expect this to be an overestimate of typical CCSN extinction for 
two reasons: (i) Balmer line production is weighted towards higher luminosity stars, and therefore younger phases of star clusters ($\sim1$--5\,Myr, G10); (ii) SNe occur at the end of a star's life, 
counts are weighted by number in the IMF, and therefore SNe typically occur
later in the life of star clusters ($\sim 10$--40\,Myr).
While we still expect some increase of extinction with host galaxy mass due to the inter-stellar 
medium, we assume a constant extinction with mass for the remaining results of the present work. 

\subsection{Star-Formation Rate Density} \label{sec:SFRD}

\begin{figure*}
	\centerline{\includegraphics[width=0.7\textwidth]{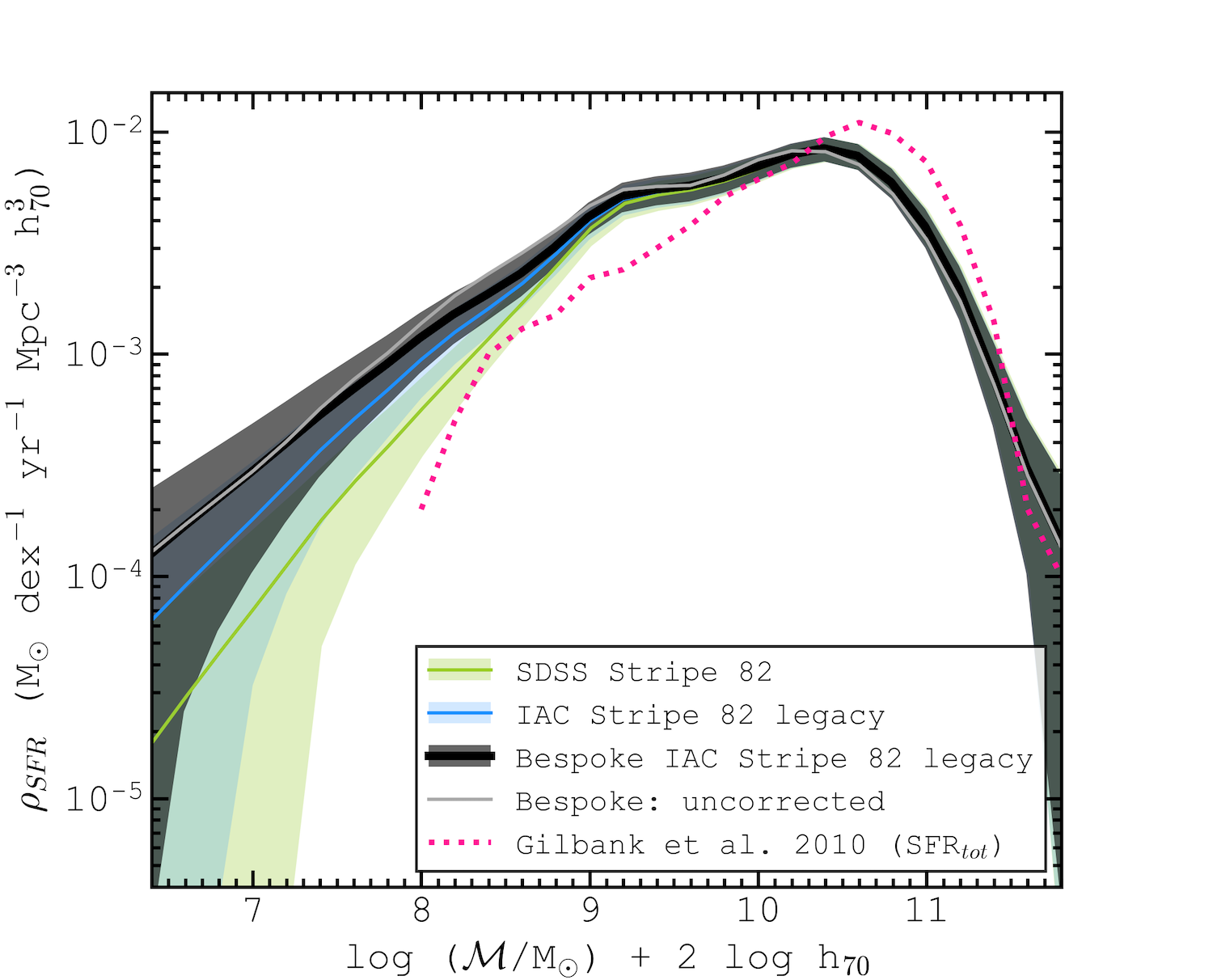}}
    \caption{\new{Volumetric} z $<$ 0.2 star-formation rate densities as a function of host galaxy stellar mass. The black line depicts galaxies selected from a bespoke search for CCSN host galaxies in IAC Stripe 82 Legacy imaging. The blue line shows galaxies from the IAC Stripe 82 legacy galaxy catalogue. The green line shows galaxies from the SDSS Stripe 82 galaxy catalogue. Shaded regions indicate 1$\sigma$ of standard deviation from Monte Carlo, Poisson and cosmic variance errors. The grey line depicts the same as the black solid line but uncorrected for SN detection efficiencies. The magenta dotted line indicates G10 \textquotesingle SFR$_{tot}$\textquotesingle\; star-formation rate densities.}
    \label{fig:rho_SFR}
\end{figure*}

To convert CCSN-rate densities into star-formation rate densities, we require an assumption for the CCSN rate per unit of star formation, $\mathcal{R}$, as discussed in Section \ref{sec:eqsbreakdown}. Using a value of log $\mathcal{R}$ = -- 1.9, 
we obtain the star-formation-rate densities plotted in Figure \ref{fig:rho_SFR}, as a function of galaxy stellar mass. Our SFRDs are consistent with the G10 SFRDs (SFR$_{tot}$) between 9.0 $<$ log($\mass$/M$_{\odot}$) $<$ 11.0, the stellar mass range for which the G10 galaxy sample is expected to be complete.

Using our bespoke search for the CCSN hosts in IAC Stripe 82 legacy imaging, results are sufficiently constrained to deduce a power-law decrease to star-formation rate densities with decreasing galaxy stellar mass, down to the low mass limit. Our method allows for an estimation of star-formation rates 1.6 dex lower in stellar mass than achieved by G10 who calculated star formation rates directly from galaxy emission lines.

\subsection{The Galaxy Stellar Mass Function}\label{sec:GSMF}

\new{For the previous results of the present work, no assumptions are required for the volumetric numbers of galaxies at each stellar mass. On the contrary, by making assumptions about specific star formation rate levels with mass, volumetric galaxy number densities as a function of mass can be estimated.} 

The star-forming GSMF can be derived from CCSN-rate densities using Equation \ref{eq:CCSNGSMF}, requiring an assumption for mean specific CCSN-rate variation with galaxy stellar mass ($\mass$). Specific CCSN-rates are expected to trace $\ssfr$ irrespective of $\mass$ (See Section \ref{sec:intro}). CCSN-rates may in fact be the most direct tracers of star formation rates. Studies which attempt to measure both $\ssfr$ and specific CCSN rates find consistent slopes with $\mass$ \cite[see, e.g.][]{GBM15}. This gives confidence in our assumption of a constant $\mathcal{R}$ value with $\mass$, as assumed to estimate star-formation rate densities in Section \ref{sec:SFRD}. With this similar trend in mind we can use the more numerous studies of $\ssfr$ vs $\mass$ to suggest the sensible range of assumptions for specific CCSN-rate vs mass, required to derive the star-forming GSMF.

There exist conflicting results in the literature for the variation of $\ssfr$ vs $\mass$. Whilst some studies find much higher efficiencies towards lower masses \citep{ZHE07,LI11,KAR11,GBM15}, where typically $\ssfr \propto \mass^{-0.5}$, others find much shallower trends consistent with a constant $\ssfr$ \citep[see, e.g.][and notably, S18]{JAM08,BEL07,WUY11}.

The majority of these studies do not probe down to the masses studied in the present work. Uncertainty exists around whether the $\ssfr$ vs $\mass$ relations found for massive galaxies apply to the dwarf regime down to {log($\mass$/M$_{\odot}$)=6.4}. \citet{MCG17} find a distinct star forming main sequence for $7 \lesssim \log(\mass$/M$_{\odot}) \lesssim 10.0$, with results consistent with a constant $\ssfr$. \new{Rate simulations} of \citet{GBM15} predict a tanh-like function to specific CCSN-rate with galaxy mass, with constant specific rates in the dwarf regime and at the highest galaxy masses (>10$^{11}$), and with decreasing $\ssfr$ with increasing mass for the masses in between.

In the present work we are most interested in the shape of the galaxy stellar mass function in the dwarf regime. High mass ($\log$($\mass$/M$_{\odot}$) $\gtrsim 9.0$) galaxy counts in this work are subject to uncertainties related to the modelling of host galaxy extinction in massive galaxies, and the star-forming GSMF is already well constrained at high masses by several independent works finding consistent number densities \citep[e.g.][]{BAL08,LW09,KEL14b}. Consider instead the mass range of $8.0 < \log(\mass$/M$_{\odot}) < 9.0$. The variation of SF galaxy counts with mass in this range is also well constrained, yet the effects of varying assumptions for host galaxy extinction with mass are significantly smaller \new{than at higher masses}. Hence we would expect that the $\ssfr$ vs $\mass$ relation which produces a slope consistent with well-constrained number densities from previous studies is the most reliable relation. 

We find the best fit $\ssfr$ vs $\mass$ log-log relation for {$8.0 < \log(\mass$/M$_{\odot}) < 9.0$} to B12 star-forming galaxy number densities (computed from GAMA data, \citealt{DRI11}) to have a slope of $-0.08 ^{+0.08}_{-0.10}$. We can express Equation \ref{eq:CCSNGSMF} in the following form:
\begin{equation}\label{eq:gamma}
\phi_{SF} \propto \rho_{CCSN} \mass^{-\gamma}  \mbox{~~~.}
\end{equation}

\begin{figure}
	\centerline{\includegraphics[width=1.22\columnwidth]{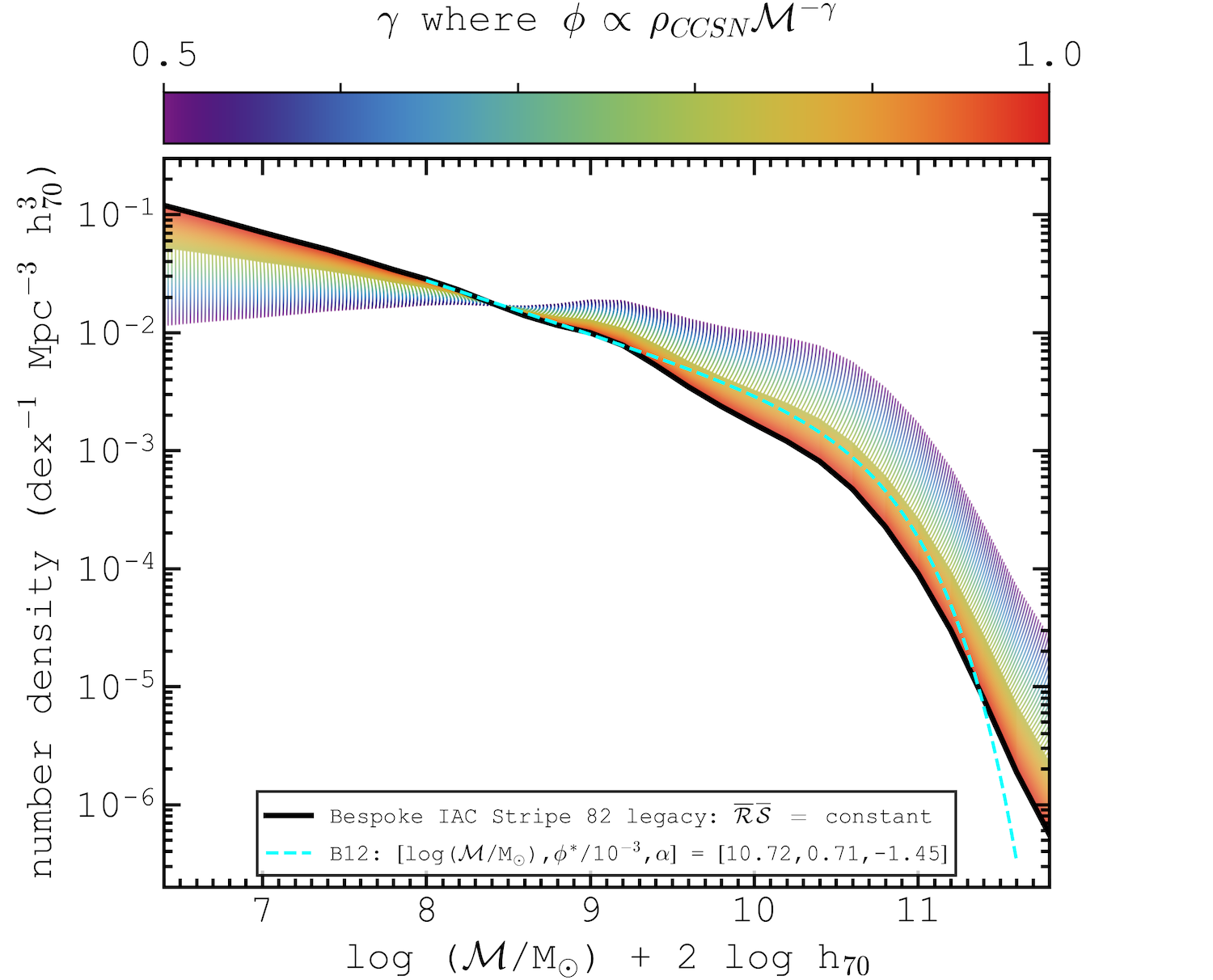}}
    \caption{The z < 0.2 star-forming galaxy stellar mass function, as a function of the parameter $\gamma(\mathcal{R},\ssfr)$. The black line shows the GSMF derived assuming constant specific CCSN-rate with stellar mass, i.e. $\gamma=1.0$. The solid region shows star-forming galaxy number densities corresponding to the 1$\sigma$ uncertainty level on the best fit value of $\gamma$ to the B12 star-forming GSMF (the cyan dashed line) in the range 8.0 < log($\mass$/M$_{\odot}$) < 9.0. The hatched region shows number densities derived using lower values of $\gamma$.}
    \label{fig:GSMF_gamma}
\end{figure}

For a constant ratio of CCSN-rate to SF-rate with mass, our best-fit $\ssfr$ vs $\mass$ slope leads to $\gamma = 0.92 ^{+0.08}_{-0.10}$. We find a constant value of $\gamma$ with mass is able to give good agreement with B12 star-forming galaxy number densities, across their full range of masses. We show the effects of different values of $\gamma$ on the derived star-forming galaxy stellar mass function in Figure \ref{fig:GSMF_gamma}. This shows the typical range for $\ssfr$ vs $\mass$ found in the literature.

\begin{figure*}
	\includegraphics[width=0.7\textwidth]{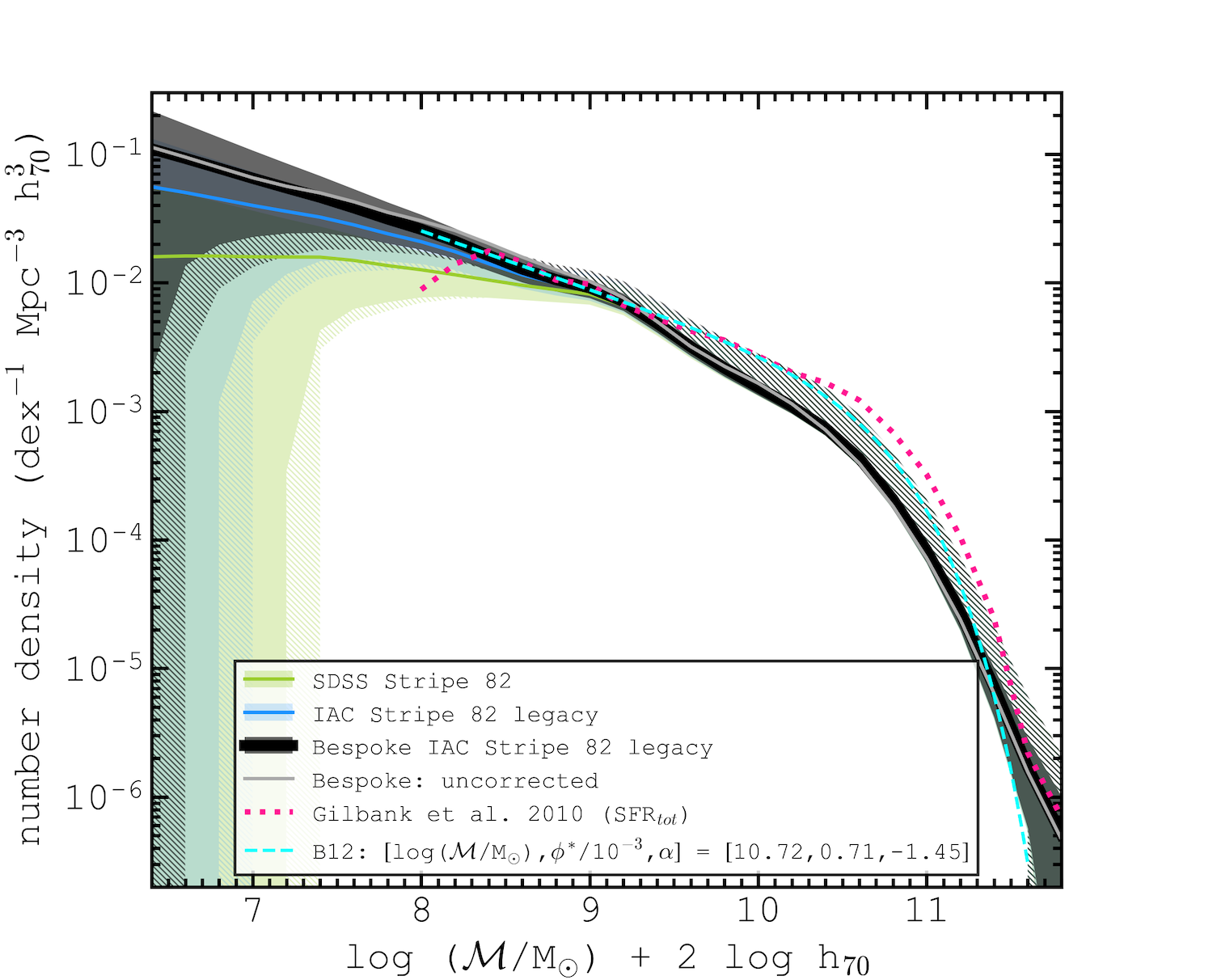}
    \caption{The z < 0.2 star-forming galaxy stellar mass function: star-forming galaxy number densities as a function of stellar mass, in logarithmic units of solar mass, as derived from corresponding star-formation rate densities (see Figure \ref{fig:rho_SFR}). The cyan dashed line represents the Schechter-fit to star-forming galaxy number densities obtained by B12. Hatched regions represent additional uncertainties on top of observational uncertainties concerning the optimal model of \new{specific star formation rate} vs galaxy mass (see Figure \ref{fig:GSMF_gamma}).}
    \label{fig:GSMF}
\end{figure*}

Given our focus is on low-mass galaxies, we assume constant $\ssfr$ with mass ($\gamma=1.0$), and normalise number densities to those of B12 in the mass range {8.0 < log($\mass$/M$_{\odot}$) < 9.0}, to derive the star-forming galaxy number densities (Mpc$^{-3}$ dex$^{-1}$) for {6.4 < log($\mass$/M$_{\odot}$) < 11.8}. This is shown in Figure \ref{fig:GSMF} with the Monte Carlo, Poisson and cosmic variance uncertainties. We once more show the results based on the 3 galaxy samples derived using SDSS Stripe 82 galaxies, IAC Stripe 82 legacy galaxies, and those from our bespoke SN host search of IAC Stripe 82 legacy imaging.

Using the full sample of CCSN host galaxies, we observe a continuation of a power-law rise \new{in galaxy number density} with decreasing mass. When selecting host galaxies from the IAC and SDSS catalogues, incompleteness is found below masses of log($\mass$/M$_{\odot}$) $\sim$ 9.0, and zero number densities cannot be ruled out below masses of log($\mass$/M$_{\odot}$) = 6.8 and log($\mass$/M$_{\odot}$) = 7.2, respectively.

We find that our result for the full sample is consistent with the Schechter function fit to B12 star-forming galaxy number densities, when extrapolating the fit below log($\mass$/M$_{\odot}$) = 8.0. B12 use a fit with parameters [log ($\mass$/M$_{\odot}$),\, $\phi^{*}/10^{-3}$ dex$^{-1}$Mpc$^{-3}$,\, $\alpha$] = [10.72,\, 0.71,\, --1.45], which is plotted in Figure \ref{fig:GSMF}. For our method, using $\gamma=1.0$, we find the best-fit parameters to be [10.54,\, 1.32,\, --1.41], obtained using a Levenberg-Marquardt algorithm applied to the full mass range shown in Figure \ref{fig:GSMF}.

We also compare with the GSMF derived from G10\textquotesingle s SFRD assuming constant specific star-formation rate with galaxy mass. Between 8.4 < log($\mass$/M$_{\odot}$) < 11.0 the gradient of number densities with mass is consistent with the B12 equivalent, giving support to this assumption.

Although we show the star-forming galaxy stellar mass function rather than that of all galaxies, it is expected that the low-mass population is dominated by star-forming galaxies (B12). Therefore, by constraining the low mass end of the star-forming GSMF we are putting strong constraints on the form of the total GSMF. We find our low-mass number densities to be consistent with those from the EAGLE simulations \citep{SCH15} assuming a standard $\Lambda$-CDM cosmology. \new{Given that a GSMF with incompleteness in the dwarf regime could be mistaken as evidence for 
tension with the standard cosmology, it is clear that our overcoming of surface brightness and stellar mass biases is crucial for an assessment of the sub-structure problem}. Consistency is also found with the low-mass galaxy number densities of \citet{WRI17}, derived from a method used to estimate and correct for surface brightness incompleteness. We find an upper limit of {0.1 dex$^{-1}$Mpc$^{-3}$} at {$10^7$ M$_{\odot}$} that is on the low end of their results, which do not distinguish between star-forming and quenched galaxies.

\section{Summary \& Conclusions}

Using a SNe sample \citep{S18} as pointers to their host galaxies, approximately 150 newly identified LSBGs have been located in IAC Stripe 82 legacy coadded imaging (Figure~\ref{fig:postagestamps}). 
This results in a significant improvement to magnitude depth of a CCSN-selected galaxy sample (Figure~\ref{fig:depth}). A galaxy selection using a complete sample of CCSNe removes surface-brightness bias. 

In order to estimate stellar masses of host galaxies without spectroscopic redshifts, 
we use a colour-based code, called zMedIC, that uses the strong correlations between Sloan \textit{ugriz} colours and redshift. 
The parameterisation used is shown to work for a sample containing both star-forming and quiescent galaxies at z < 0.4 (Figure~\ref{fig:zMedIC}). In order to assess uncertainties on CCSN-rate densities as a function of galaxy stellar mass, we employ a Monte Carlo method to fold in errors on photometric redshift
(Figure~\ref{fig:PDFs}) and $ugriz$ photometry. 
The observed CCSN-rate densities as a function of mass are shown to peak at 
$\sim 10^{10.5}\mathrm{M}_{\odot}$ and drop by a factor of $\sim100$ down to $10^{6.5}\mathrm{M}_{\odot}$ 
(Figure~\ref{fig:rho_CCSN_uncorr}).

We use a model to correct CCSN-rate densities for SN detection efficiencies ($\epsilon$) that depend on host galaxy extinction, Galactic extinction, SN type \citep{RIC14} and redshift. 
Corrected CCSN-rate densities are consistent with expectations derived from the cosmic star-formation history \citep{MD14} assuming $\log \mathcal{R} \simeq -2$ (Figure~\ref{fig:rho_CCSN_z}), where $\mathcal{R}$ is the expected number of stars that explode as CCSNe per solar mass of stars formed.

By assuming a value for $\mathcal{R}$, the corrected CCSN-rate density as a function of stellar mass
(Figure~\ref{fig:rho_CCSN_corr}) 
can then be scaled to the SFRD (Figure~\ref{fig:rho_SFR}). 
The SFRD is consistent with the emission-line derived SFRD at high masses \citep{G10} but our method extends the
measurement to lower masses. 
\new{By} assuming a constant specific star formation rate ($\ssfr$), and scaling appropriately (to \citealt{B12}), we convert the SFRD to the star-forming GSMF (Figure~\ref{fig:GSMF}). 
Derived star-forming galaxy number densities are found to rise as a power-law with decreasing galaxy stellar mass down to the lowest assessed masses of log($\mass$/M$_{\odot}$) = 6.4;
and even at the lower end of our estimated uncertainties, 
there is no turnover in the number density of star-forming galaxies per unit log mass at least down to log($\mass$/M$_{\odot}) \sim 7$. 

We have demonstrated a method which significantly reduces tension between observations and the simulated predictions of galaxy number densities derived via a standard $\Lambda$-CDM cosmology.
The lower \new{detection} limit to galaxy stellar mass for the SN method outlined in this paper depends on the area, depth and the observing period of the SN survey. 
A future sample derived from LSST time-series and coadded imaging could significantly increase
the number of reliably identified CCSN hosts at z < 0.2. 
As part of this, deep multi-band photometry is crucial for constraining photometric redshifts and 
stellar masses of LSBGs.
This would enable measurements to even lower masses, and future work would also
enable a more detailed assessment of the functional form of 
$\mathcal{R}$, $\epsilon$ and $\ssfr$ on the conversion from observed CCSN-rate density to SFRD 
and then to GSMF. 

\section{Acknowledgments}

Funding for the SDSS-II, SDSS-III and SDSS-IV has been provided by the Alfred P. Sloan Foundation, the National Science Foundation, the U.S. Department of Energy Office of Science, the National Aeronautics and Space Administration, the Japanese Monbukagakusho, the Max Planck Society, the Higher Education Funding Council for England, and the Participating Institutions. SDSS-IV acknowledges support and resources from the Center for High-Performance Computing at the University of Utah. The SDSS web site is www.sdss.org.

SDSS-II, SDSS-III and SDSS-IV are managed by the Astrophysical Research Consortium for the 
Participating Institutions of the SDSS Collaboration including 
the American Museum of Natural History,
University of Arizona, 
Astrophysical Institute Potsdam, 
University of Basel, 
the Brazilian Participation Group, 
Brookhaven National Laboratory, 
University of Cambridge, 
the Carnegie Institution for Science, 
Carnegie Mellon University, 
Case Western Reserve University, 
University of Chicago, 
the Chilean Participation Group,
University of Colorado Boulder,
Drexel University, 
Fermilab, 
University of Florida, 
the French Participation Group,
the German Participation Group,
Harvard-Smithsonian Center for Astrophysics,
Harvard University, 
the Institute for Advanced Study, 
Instituto de Astrof\'isica de Canarias, 
the Japan Participation Group, 
Johns Hopkins University, 
the Joint Institute for Nuclear Astrophysics, 
the Kavli Institute for Particle Astrophysics and Cosmology, 
Kavli Institute for the Physics and Mathematics of the Universe (IPMU) / University of Tokyo, 
the Korean Scientist Group, 
the Chinese Academy of Sciences (LAMOST), 
Lawrence Berkeley National Laboratory, 
Leibniz Institut f\"ur Astrophysik Potsdam (AIP),  
Los Alamos National Laboratory, 
Max-Planck-Institut f\"ur Astronomie (MPIA Heidelberg), 
Max-Planck-Institut f\"ur Astrophysik (MPA Garching), 
Max-Planck-Institut f\"ur Extraterrestrische Physik (MPE), 
the Michigan State/Notre Dame/JINA Participation Group, 
Universidad Nacional Aut\'onoma de M\'exico, 
National Astronomical Observatories of China,
New Mexico State University, 
New York University, 
University of Notre Dame, 
Observat\'ario Nacional / MCTI,
Ohio State University, 
University of Oxford, 
Pennsylvania State University, 
University of Pittsburgh, 
University of Portsmouth, 
Princeton University,
Shanghai Astronomical Observatory, 
the Spanish Participation Group, 
University of Tokyo, 
the United States Naval Observatory, 
University of Utah, 
Vanderbilt University, 
United Kingdom Participation Group,
University of Virginia, 
the University of Washington,
University of Wisconsin, 
Vanderbilt University, 
and Yale University.

\bibliography{main.bbl}

\label{lastpage}
\end{document}